\documentclass[aps,prb,superscriptaddress,amsfonts,amsmath,amssymb,showpacs,floatfix,reprint,longbibliography,footinbib]{revtex4-2}
\usepackage{url}
\usepackage{bm}
\usepackage{graphicx}
\usepackage{amsmath}
\usepackage{amstext}
\usepackage{amssymb}
\usepackage{amsfonts}
\usepackage{amsbsy}
\usepackage{verbatim}
\usepackage{color}
\usepackage[colorlinks=true, urlcolor=blue, linkcolor=blue, citecolor=blue, pdftex]{hyperref}
\usepackage[capitalise]{cleveref}
\usepackage{multirow}
\usepackage{floatrow}
\usepackage{float}
\usepackage{tikz}
\usepackage{gensymb}
\usepackage{textcomp}
\usepackage{enumitem}
\usepackage[version=3]{mhchem}
\usepackage{afterpage}
\usepackage{pbox}
\usepackage{makecell}
\usepackage{dsfont}
\usepackage{siunitx}
\usepackage{stackengine,wasysym,scalerel}
\usepackage{latexsym}
\usepackage{cancel}
\usepackage{array, multirow, bigdelim, makecell, booktabs}
\usepackage[misc]{ifsym}
\usepackage{times}
\usepackage{longtable}
\usepackage[title]{appendix}

\begin{document}

\title{Dzyaloshinskii–Moriya-driven instabilities in square-kagome quantum antiferromagnets}

\author{Leonid S. Taran}
\thanks{These authors contributed equally.}
\affiliation{M. N. Mikheev Institute of Metal Physics, Ural Branch of Russian Academy of Sciences, 620137 Ekaterinburg, Russia}
\author{Arnaud Ralko}
\thanks{These authors contributed equally.}
\affiliation{Institut Néel, Université Grenoble Alpes, CNRS, Grenoble INP, 38000 Grenoble, France}
\affiliation{Department of Physics, Indian Institute of Technology Madras, Chennai 600036, India}
\author{Fedor V. Temnikov}
\thanks{These authors contributed equally.}
\affiliation{M. N. Mikheev Institute of Metal Physics, Ural Branch of Russian Academy of Sciences, 620137 Ekaterinburg, Russia}
\author{Vladimir V. Mazurenko}
\affiliation{Department of Theoretical Physics and Applied Mathematics, Ural Federal University, Mira St. 19, 620002 Ekaterinburg, Russia}
\author{Sergey V. Streltsov}
\affiliation{M. N. Mikheev Institute of Metal Physics, Ural Branch of Russian Academy of Sciences, 620137 Ekaterinburg, Russia}
\affiliation{Department of Physics, Indian Institute of Technology Madras, Chennai 600036, India}
\affiliation{Department of Theoretical Physics and Applied Mathematics, Ural Federal University, Mira St. 19, 620002 Ekaterinburg, Russia}
\author{Yasir Iqbal}
\email{yiqbal@physics.iitm.ac.in}
\affiliation{Department of Physics, Indian Institute of Technology Madras, Chennai 600036, India}

  \date{\today}

\begin{abstract}
Decorated square-kagome quantum antiferromagnets provide a natural setting in which strong frustration, lattice decoration, and spin--orbit-induced anisotropy compete on comparable energy scales. Here we show that in Na$_6$Cu$_7$BiO$_4$(PO$_4$)$_4$Cl$_3$ the coupling ($J_{10}$) which links the decorating Cu(3) sites to the square-kagome backbone, stabilizes the gapped quantum-paramagnetic regime, while symmetry-allowed Dzyaloshinskii--Moriya (DM) interactions systematically suppress the minimum spinon gap $\Delta_{\mathrm{spinon}}$ and drive the system toward magnetic condensation. To establish this, we combine \emph{ab initio} calculation of the DM vectors with a generalized Schwinger-boson self-consistent mean-field theory that treats singlet and triplet hopping/pairing channels on equal footing. As a benchmark, the isotropic square-kagome Heisenberg model exhibits four competing low-energy saddle points distinguished by their Wilson-loop fluxes and by characteristic static and dynamical structure-factor fingerprints. A minimal DM perturbation does not qualitatively reshape this competing landscape, but already enhances the tendency towards order. For the realistic decorated Hamiltonian, finite-size scaling of $\Delta_{\mathrm{spinon}}$ together with momentum-resolved structure factors identifies $J_{10}$ (exchange with decorating Cu) as the control parameter of the gapped regime and shows that the full symmetry-allowed DM pattern shifts the system further toward condensation. Our results place Na$_6$Cu$_7$BiO$_4$(PO$_4$)$_4$Cl$_3$ in close proximity to a magnetic instability and provide experimentally testable predictions for anisotropy-enhanced soft modes in decorated square-kagome materials.
\end{abstract}

\maketitle

\section{Introduction}

Frustrated quantum magnets built from corner-sharing motifs continue to provide a fertile setting for unconventional collective behavior, including dense low-energy spectra, strong renormalization of thermodynamic scales, and quantum-disordered ground states. Within this broader landscape, the square-kagome (shuriken) lattice occupies a distinctive position~\cite{Siddharthan-2001}. Like kagome, it combines strong geometric frustration with corner-sharing triangles, but it also hosts square plaquettes, flat-band physics, and a different hierarchy of competing ordered and disordered states~\cite{Rousochatzakis-2013,Jahromi2025,ralko_resonating-valence-bond_2015,merino_role_2018,Yogendra-2024}. As a result, the spin-$\tfrac{1}{2}$ square-kagome Heisenberg antiferromagnet has emerged as a nontrivial frustrated platform in its own right, with high-field anomalies~\cite{Derzhko2014,Hasegawa-2018,Morita2018}, low-temperature thermodynamic signatures~\cite{Tomczak-2003,Richter2022}, and competing valence-bond and ordered tendencies already established theoretically \cite{Nakano-2013,Richter2023,Gembe-2023,Schmoll-2023,Jahromi2025}.

This theoretical interest has been amplified by the recent emergence of square-kagome and decorated square-kagome materials. Hydrothermal synthesis and structural characterization of Na$_6$Cu$_7$BiO$_4$(PO$_4$)$_4$[Cl,(OH)]$_3$ established a prominent Cu$^{2+}$-based realization in which a square-kagome layer is decorated by an additional magnetic Cu site \cite{Yakubovich2021,Popova1987,Pertlik1988}. Low-temperature thermodynamic measurements on Na$_6$Cu$_7$BiO$_4$(PO$_4$)$_4$[Cl,(OH)]$_3$ and related compounds revealed the absence of conventional long-range order down to very low temperatures together with a pronounced low-energy sector \cite{Liu2022,Shvanskaya-2026}. At the same time, the broader nabokoite family and related tellurate--sulfate compounds~\cite{Murtazoev2023,Markina2022,Fujihala-2020,Rebrov-2024,Goto-2024,shvanskaya2026} have expanded the experimental landscape of decorated and distorted square-kagome magnets, making it increasingly necessary to go beyond idealized isotropic nearest-neighbor models~\cite{Gonzalez-2025}. These developments elevate square-kagome magnetism from a model curiosity to a growing materials platform.

A key lesson from this materials progress is that the relevant exchange networks are not only frustrated but also structurally low-symmetry. The present work is motivated by the fact that in low-symmetry Cu--O--Cu environments relevant to decorated square-kagome materials, Dzyaloshinskii--Moriya (DM) interactions are not a secondary correction but a symmetry-allowed perturbation of experimentally relevant magnitude. The antisymmetric exchange introduced by Dzyaloshinskii and Moriya \cite{Dzyaloshinskii1958,Moriya1960} is well known to qualitatively reshape frustrated spin-$\tfrac{1}{2}$ magnets by lowering spin-rotation symmetry, admixing chiral correlations, and stabilizing magnetic order out of nearby quantum-disordered regimes. In kagome antiferromagnets, for example, DM interactions shift phase boundaries, modify excitation spectra, and enhance condensation tendencies within Schwinger-boson descriptions \cite{Messio2010,Mondal2017}. In the square-kagome setting, where competing states are already close in energy, the effect of symmetry-allowed DM interactions is therefore a central and unavoidable question.

This issue is especially acute for decorated square-kagome compounds. In our recent work on Na$_6$Cu$_7$BiO$_4$(PO$_4$)$_4$Cl$_3$, we established a realistic antiferromagnetic Hamiltonian and showed that the material remains quantum paramagnetic despite the presence of the decorating Cu$^{2+}$ site \cite{Niggemann2023}. An important outcome of that work was that the decorating site is not inert: the coupling $J_{10}$, which connects the decorating apical Cu(3) sites to the square-kagome backbone, plays an active role in stabilizing the disordered regime. This immediately raises the next microscopic question: how robust is that quantum-paramagnetic regime once symmetry-allowed anisotropic exchange is included? More specifically, does DM merely perturb the disordered state, or does it systematically drive the system toward magnetic condensation?

In this paper we show that the latter question is governed by a simple and physically transparent mechanism. In decorated square-kagome antiferromagnets, $J_{10}$ stabilizes the quantum-paramagnetic regime, while symmetry-allowed DM interactions systematically destabilize it and drive the system toward magnetic ordering. To establish this, we combine two complementary ingredients. First, we extract the full symmetry-allowed DM vectors for Na$_6$Cu$_7$BiO$_4$(PO$_4$)$_4$Cl$_3$ from \emph{ab initio} calculations incorporating spin-orbit-coupling. Second, we analyze both the ideal and material-specific Hamiltonians within a generalized Schwinger-boson self-consistent mean-field theory that treats singlet and triplet hopping/pairing channels on equal footing \cite{ArovasAuerbach1988,ReadSachdev1991,Manuel1996}. This framework allows us to track how anisotropy reshapes the spinon spectrum, shifts the proximity to boson condensation, and modifies equal-time and dynamical correlation signatures.

Our analysis proceeds in two steps. We first benchmark the generalized formalism on the isotropic square-kagome model, where we identify the competing low-energy saddle points and test the effect of a minimal DM perturbation. We then turn to the realistic compound Hamiltonian of Na$_6$Cu$_7$BiO$_4$(PO$_4$)$_4$Cl$_3$, where the central result emerges: the coupling $J_{10}$ acts as the control parameter stabilizing the gapped regime, while the symmetry-allowed DM pattern suppresses the minimum spinon gap and enhances the tendency toward ordering. This places the material in close proximity to a magnetic instability and yields experimentally testable predictions for anisotropy-enhanced soft modes and momentum-resolved spectral weight redistribution.

The remainder of the paper is organized as follows. In Sec.~\ref{dm} we extract the isotropic exchanges and symmetry-allowed DM vectors for Na$_6$Cu$_7$BiO$_4$(PO$_4$)$_4$Cl$_3$ from \emph{ab initio} calculations and summarize the resulting microscopic Hamiltonian. In Sec.~\ref{sec:sbmft} we introduce the generalized Schwinger-boson formalism, define the Wilson-loop diagnostics, and describe the numerical protocol. In Sec.~\ref{results} we present the results for the isotropic square-kagome benchmark, the minimal DM perturbation test, and the realistic compound Hamiltonian, including the evolution of the minimum spinon gap and the corresponding equal-time and dynamical structure factors. Finally, in Sec.~\ref{conclusion} we summarize the physical implications for decorated square-kagome quantum antiferromagnets and for ongoing experiments on this growing materials family.

\section{DM vectors extraction from DFT calculations}
\label{dm}

\begin{figure}[h!]
\centering
\includegraphics[width=\columnwidth]{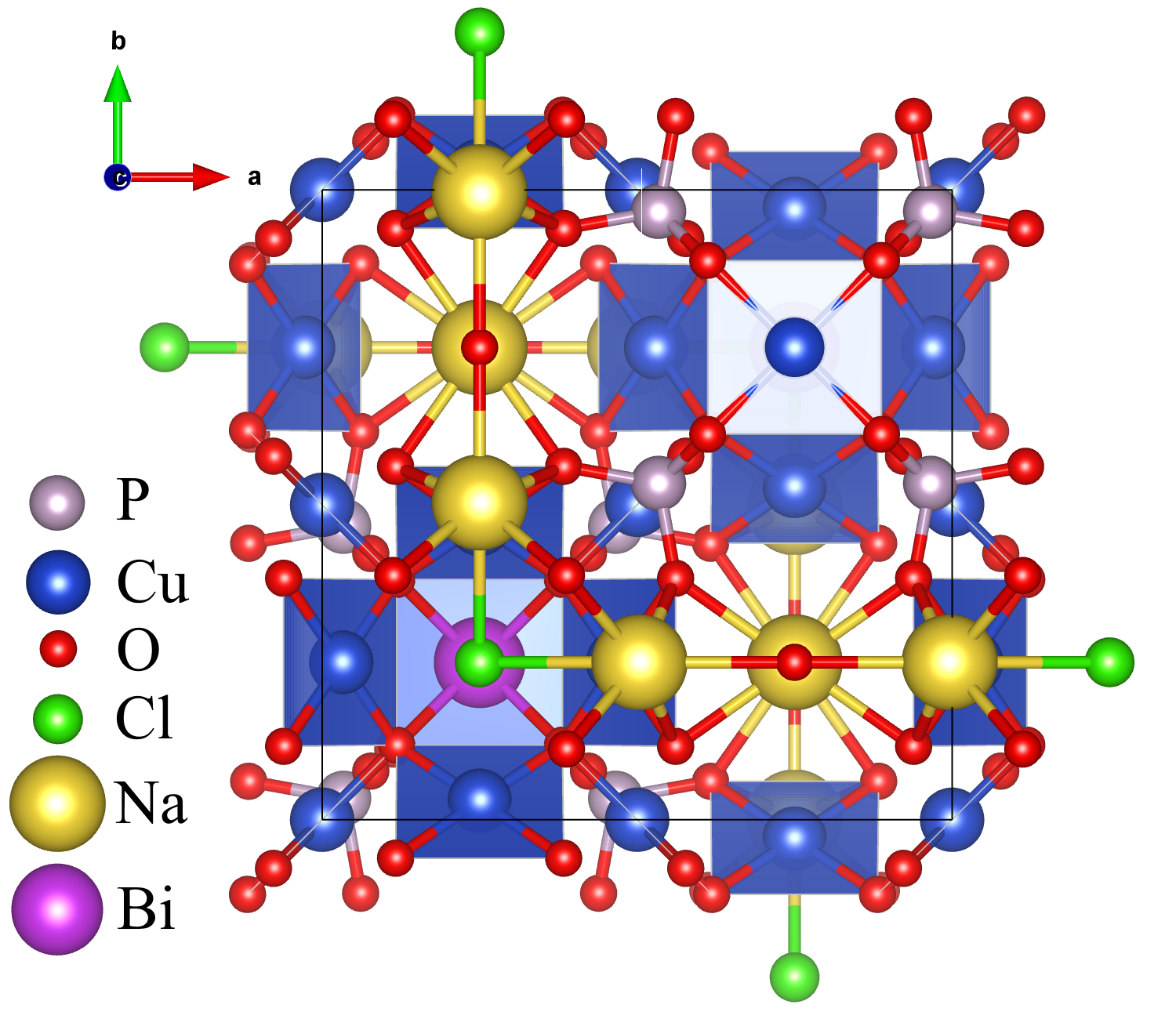}
    \caption{Crystal structure of Na$_6$Cu$_7$BiO$_4$(PO$_4$)$_4$Cl$_3$. The Cu$^{2+}$ ions occupy three crystallographically inequivalent positions, Cu(1), Cu(2), and Cu(3). Cu(1) and Cu(2) form the square-kagome backbone, while Cu(3) decorates it in a checkerboard manner above and below the square-kagome plane.}
    \label{Na6Cu7_crystal_structure}
\end{figure}

The crystallographic data were taken from Ref.~\cite{Yakubovich2021}, and the crystal structure is shown in Fig.~\ref{Na6Cu7_crystal_structure}. The conventional unit cell contains two formula units of Na$_6$Cu$_7$BiO$_4$(PO$_4$)$_4$Cl$_3$, crystallizes in the tetragonal $P4/nmm$ space group. Each Cu ion is coordinated by four oxygen ligands forming square CuO$_4$ plaquettes. Given this local coordination environment and the fact that Cu$^{2+}$ is a $d^9$ ion with spin $S = 1/2$, one expects the $x^2-y^2$ orbital to remain half-filled and thus to dominate the low-energy electronic structure.

\begin{figure}[t!]
\centering
\includegraphics[width=1\columnwidth]{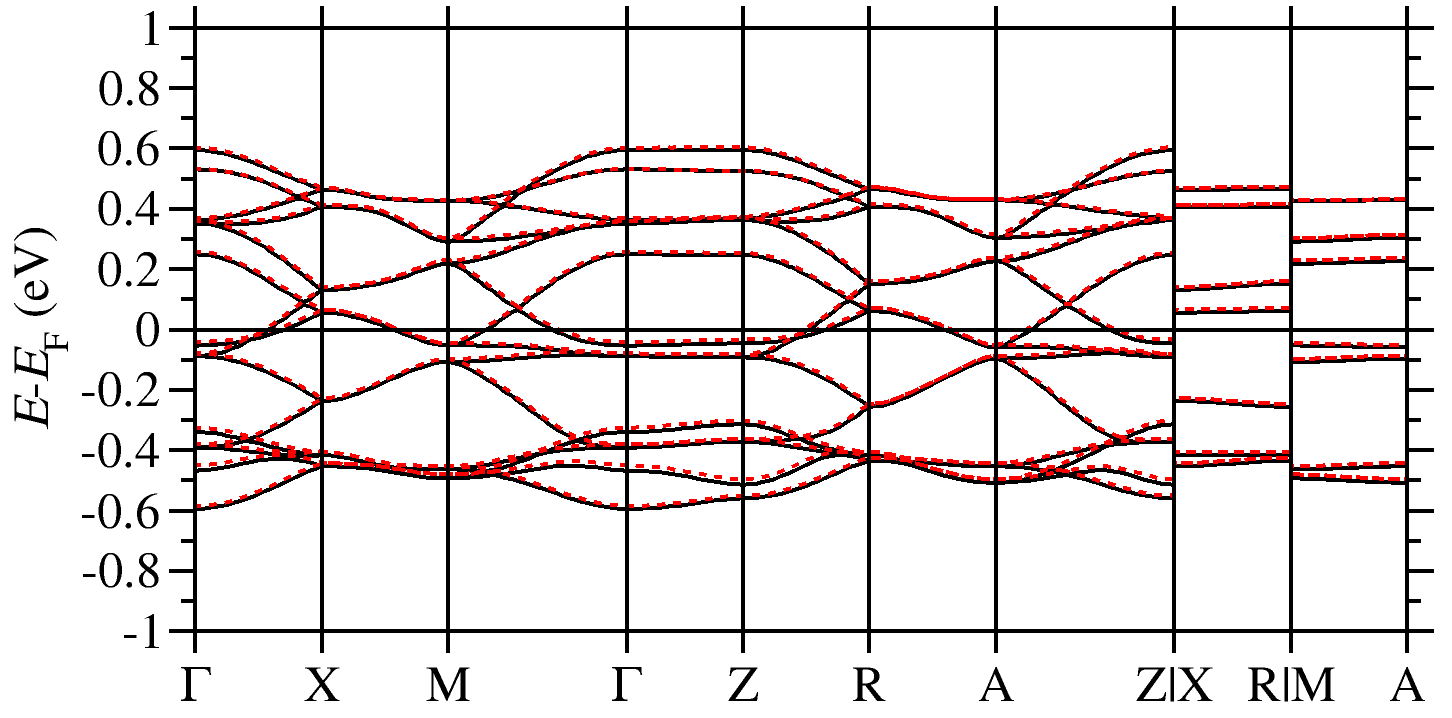}
    \caption{Band structure of Na$_6$Cu$_7$BiO$_4$(PO$_4$)$_4$Cl$_3$ from nonmagnetic GGA calculations. Dashed red lines show the DFT bands, while solid black lines show the Wannier-interpolated bands obtained from the Cu $d_{x^2-y^2}$ orbitals used to estimate the exchange parameters via hopping integrals. The isolated 14-band manifold near the Fermi level supports the one-band low-energy description.}
    \label{Na6Cu7_GGA_bands}
\end{figure}

The band structure shown in Fig.~\ref{Na6Cu7_GGA_bands} reveals a group of 14 bands near the Fermi level originating from the empty Cu $d_{x^2-y^2}$ states. These are separated from the remaining valence $d$ states by an energy gap of 400~meV. Further details of the electronic structure are provided in the form of the density of states in Fig.~\ref{Na6Cu7_GGA_DOS} of Appendix~\ref{appendix_A}.

\begin{figure}[t!]
\centering
\includegraphics[width=0.9\columnwidth]{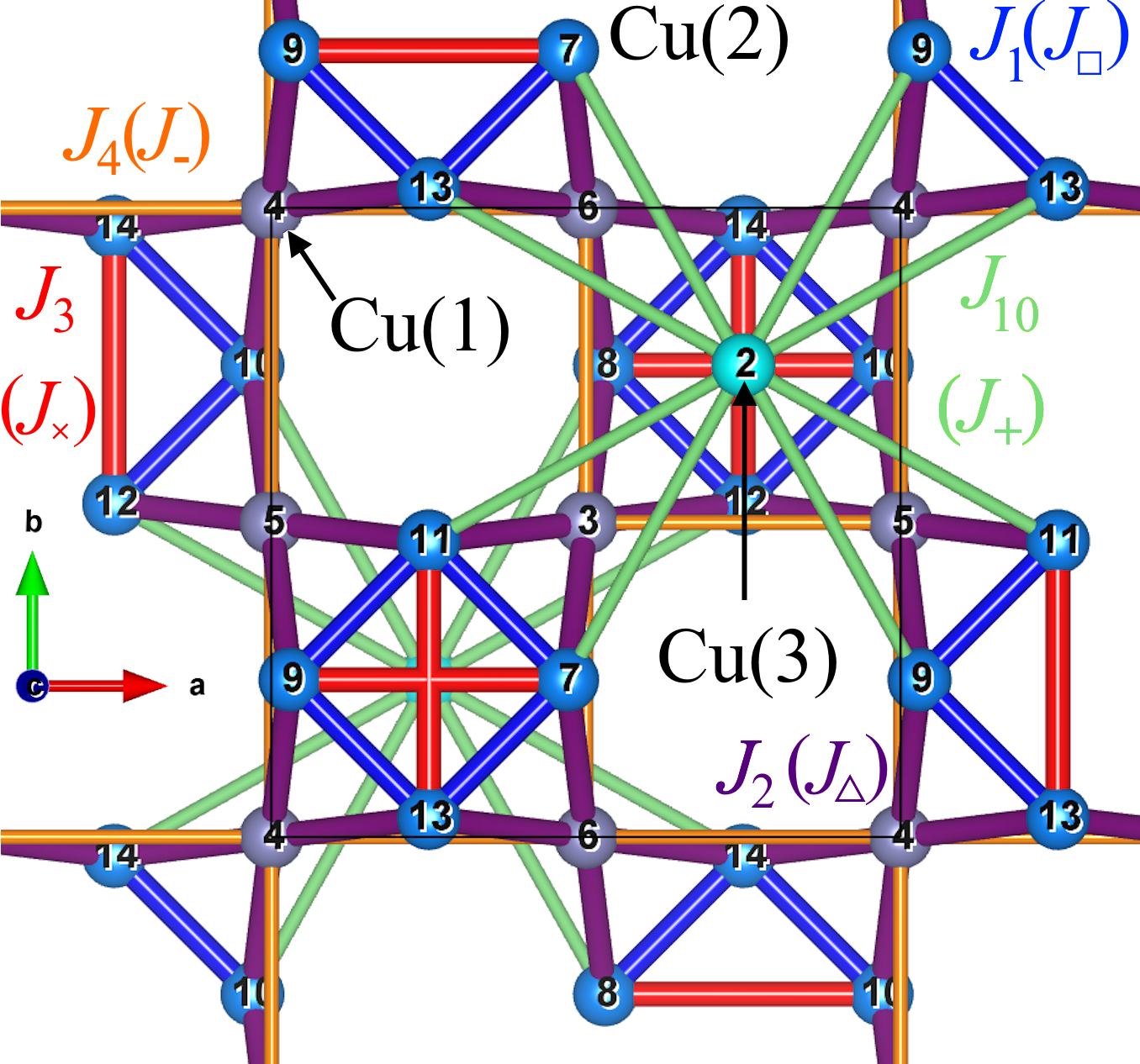}
\caption{Relevant exchange paths in Na$_6$Cu$_7$BiO$_4$(PO$_4$)$_4$Cl$_3$ with Cu sites labeled for clarity. Cu(1) and Cu(2) form the square-kagome backbone, while Cu(3) denotes the decorating sites. Colored bonds distinguish the different isotropic exchange couplings entering the microscopic spin Hamiltonian.}
    \label{Na6Cu7_exchanges}
\end{figure}

Within the unit cell, the Cu$^{2+}$ ions occupy three crystallographically inequivalent positions, Cu(1), Cu(2), and Cu(3), indicated by different colors in Fig.~\ref{Na6Cu7_exchanges}. Cu(1) and Cu(2) form a square kagome lattice, while Cu(3) decorates it at the top and bottom in a checkerboard manner. The latter are not directly connected to other Cu atoms through shared oxygen ligands, making them isolated from the kagome lattice. These atoms form square plaquettes with a uniform Cu--O bond length of 1.959~\AA.

The Cu(1) plaquettes exhibit elongation and adopt a rhombic shape. Here, the longer Cu--O bonds (1.952~\AA) connect Cu(1) with Cu(2) atoms, while the shorter bonds (1.938~\AA) link to the [PO$_4$] tetrahedra.

The Cu(2) atoms not only interconnect with each other and Cu(1), but are also attracted to bismuth atoms located above or below the center of the copper squares. This leads to a tilting of the CuO$_4$ plaquettes and results in a trapezoidal distortion with two distinct Cu--O bond lengths: a shorter pair (1.929~\AA) and a longer pair (1.964~\AA).

To estimate the parameters of exchange interaction, maximally localized Wannier functions corresponding to the Cu $x^2-y^2$ orbitals were constructed. The distinct geometries of the plaquettes influence the hybridization and the band structure, and therefore also affect the spatial localization of the Wannier functions. As a result, the smallest WF spread, approximately $\sim$2.9~\AA, is observed for the isolated plaquettes formed by the decorating Cu(3) atoms, while the largest one, $\sim$3.9~\AA, appears for the trapezoidal plaquettes of the Cu(2) sites.

The isotropic exchange interactions were estimated using Anderson's superexchange theory \cite{Anderson1959, Yildrim1995,khomskii2024}:
\begin{align}
J_{ij} = \frac{4t_{ij}^2}{U},
\label{formula: J}
\end{align}
where $t_{ij}$ are the hopping integrals and $U$ is the effective on-site Coulomb interaction for the one-band model. In Table~\ref{table:J_for_Na6Cu7_1} we present exchange couplings with $|J|>4$ K (the full list of exchanges is presented in Table \ref{table:J_for_Na6Cu7_3} of Appendix~\ref{appendix_B}).

For a reliable set of exchange parameters, good agreement between the theoretical and experimental Curie--Weiss temperatures is essential. Using the formula
\begin{equation}
    \Theta = -\frac{S(S+1)}{3} \sum_{i\neq j} J_{ij},
\label{formula: Theta_CW}
\end{equation}
where the sum runs over all pairs of copper atoms within the unit cell using the exchange couplings from Table~\ref{table:J_for_Na6Cu7_1}, one finds $\Theta = -216\,\mathrm{K}$, $-205\,\mathrm{K}$, and $-180\,\mathrm{K}$ for $U = 6.66$, $7$, and $8\,\mathrm{eV}$, respectively. These values can be compared with the experimentally measured Curie--Weiss temperature $\Theta_{\mathrm{exp}} = -212\,\mathrm{K}$. The closest agreement is achieved for $U = 6.66\,\mathrm{eV}$, as in Ref.~\cite{Niggemann2023}, where $J_{ij}$ were calculated by the accurate total energy methods, containing contribution from all possible exchange channels.  This value of $U = 6.66\,\mathrm{eV}$ is therefore adopted for all subsequent exchange-parameter calculations in $\mathrm{Na_6Cu_7BiO_4(PO_4)_4Cl_3}$.

\begin{table}[b!]
    \centering
    \caption{Isotropic exchange interactions $J_N$ and their notation $J_{\mathrm{not.}}$ (in K) for Na$_6$Cu$_7$BiO$_4$(PO$_4$)$_4$Cl$_3$, estimated from Eq.~(\ref{formula: J}) in the present work and compared with the DFT energy-mapping results of Niggemann \textit{et al.}~\cite{Niggemann2023}. The exchange-path notation is defined in Fig.~\ref{Na6Cu7_exchanges}. Only couplings with $|J|>4$ K are listed here; the full set is given in Table~\ref{table:J_for_Na6Cu7_3}.}
    \begin{ruledtabular}
    \begin{tabular}{cccccc}
        $J_{N} (J_{not.})$ & $d$ (\AA) & $U$=6.66 eV & $U$=7 eV& $U$=8 eV & Ref.\cite{Niggemann2023} \\ \midrule
        $J_1 (J_\square)$ & 3.114 & 157.7 & 150.0 & 131.3 & 109.2 \\ 
        $J_2(J_\triangle)$ & 3.278 & 202.8 & 193.0 & 168.8 & 186.3 \\
        $J_3(J_\times)$ & 4.404 & 87.3 & 83.0 & 72.7 & 155.3 \\
        $J_4(J_-)$ & 5.009 & 29.4 & 27.9 & 24.4 & 47.0 \\
        $J_5$ & 5.009 & 9.0 & 8.6 & 7.5 & 5.3 \\
        $J_{10}(J_+)$ & 6.018 & 47.8 & 45.5 & 39.8 & 64.6 \\
        $J_{14}$ & 7.084 & 5.0 & 4.8 & 4.2 & 1.9
    \end{tabular}
    \end{ruledtabular}
    \label{table:J_for_Na6Cu7_1}
\end{table}

The extracted exchange hierarchy can be discussed in the light of the square-kagome phase diagram. According to Ref.~\cite{Richter2023,Jahromi2025}, which considers only nearest-neighbor ($J_1$) and next-nearest-neighbor ($J_2$) interactions, a non-magnetic singlet ground state with a finite spin gap separating singlet and triplet excitations is expected when $0.77 < J_2/J_1 < 1.65$. For $J_2/J_1 \gtrsim 1.65$, the system adopts a ferrimagnetic ground state. A similar ratio was reported in Ref.~\cite{Niggemann2023}, with $J_2/J_1 \sim 1.7$. However, the presence of a substantial diagonal coupling $J_3$ ($J_3/J_1 \sim 1.45$) increases frustration and stabilizes a non-magnetic valence-bond crystal (VBC) ground state with broken translational symmetry.

In the present calculations, the ratio $J_2/J_1 \simeq 1.29$ clearly falls within the regime favoring a singlet ground state. This conclusion remains valid even if one considers a smaller contribution coming from $J_3$ ($J_3/J_1 = 0.55$). Furthermore, the decorating Cu(3) sites exhibit a significant exchange coupling to the main kagome network via $J_{10} = 47.8$~K.

We next turn to the antisymmetric anisotropic exchange. The DM vector $\mathbf{D}_{ij}$ is calculated using Moriya's superexchange theory \cite{Moriya1960, Yildrim1995}:
\begin{align}
\mathbf{D}_{ij} = \frac{i}{U} \left[ \text{Tr}(\hat{t}_{ij}) \text{Tr}(\hat{t}_{ji} \boldsymbol{\sigma}) - \text{Tr}(\hat{t}_{ji}) \text{Tr}(\hat{t}_{ij} \boldsymbol{\sigma}) \right],
\label{formula: DM}
\end{align}
where $\hat{t}_{ij}$ is the hopping matrix including the spin-orbit coupling, and $\boldsymbol{\sigma}$ are the Pauli matrices. This approach has been successfully applied to other compounds in Refs.~\cite{Badrtdinov2016,Badrtdinov2018}.

To evaluate the Dzyaloshinskii--Moriya interaction (DMI) parameters, the corresponding Hamiltonian from calculations including spin-orbit coupling was derived. The largest final WF spread, $\sim$3.8~\AA, occurs for the rhombic plaquettes of the Cu(1) sites. The computed leading DMI parameters, represented by one bond from each symmetry-equivalent set, are presented in Table~\ref{table:D_for_Na6Cu7_1bond}, while the full list of estimated values is summarized in Appendix~\ref{appendix_B}, Table \ref{table:D_for_Na6Cu7} and \ref{table:abs_D_for_Na6Cu7}.

The DM vectors are defined in a global coordinate system aligned with the crystallographic axes of the unit cell. Due to the antisymmetric nature of the exchange, they satisfy the relation $\mathbf{D}_{ij} = -\mathbf{D}_{ji}$. All DM vectors of a given type are related by the symmetry operations of the space group $P4/nmm$. For instance, the transformation from the nearest-neighbor Cu7--Cu11 bond to the Cu13--Cu7 bond corresponds to a $90^\circ$ rotation around the $z$ axis. This operation leaves the $z$ component of both the position vector and the Dzyaloshinskii--Moriya vector unchanged. Since the DM vector is axial, it remains invariant under inversion; therefore, the following relations must hold:
$\mathbf{D}_{\text{7--11}} = \mathbf{D}_{\text{8--12}}$,
$\mathbf{D}_{\text{13--7}} = \mathbf{D}_{\text{14--8}}$,
$\mathbf{D}_{\text{9--13}} = \mathbf{D}_{\text{10--14}}$, and
$\mathbf{D}_{\text{11--9}} = \mathbf{D}_{\text{12--10}}$.
As demonstrated in Table~\ref{table:D_for_Na6Cu7} in Appendix~\ref{appendix_B}, our calculations reproduce these relations with high accuracy.

The largest Dzyaloshinskii--Moriya interaction (DMI) parameters are found for the nearest-neighbor Cu(1)--Cu(1), Cu(1)--Cu(2), and next-nearest-neighbor Cu(1)--Cu(1) and Cu(2)--Cu(2) pairs, with magnitudes of approximately 14 K, 28 K, 9 K, and 11 K, respectively. All other DMI parameters, however, do not exceed 4 K.

The site-resolved sums of the DM vectors, $\mathbf{D}^{\text{eff}}_i = \sum_j \mathbf{D}_{ij}$, for all copper atoms are illustrated in Fig.~\ref{fig:sum_D_ab} of Appendix~\ref{appendix_B}. 
It is worth noting that the direction of the total DM vector for the Cu(2) sites lies along one of the axes within the $ab$-plane and transforms via a $90^\circ$ rotation around the $z$-axis (see Fig.~\ref{fig:sum_D_Cu(2)} of Appendix~\ref{appendix_B}).
Since the symmetry of the spin canting is defined by the symmetry of the magnetic torque, the rotation of the spins for the Cu(2) sublattice occurs within the plane perpendicular to the DM vector, i.e., in the $ac$- or $bc$-plane depending on the specific Cu(2) site orientation. 
In contrast, for the Cu(1) atoms, the total DM interaction is directed perpendicular to the Cu(1)O$_4$ plaquette planes. 
This geometry forces the spin rotation to occur strictly within the plane of the plaquettes.
Finally, for the isolated decorating Cu(3) sites, the calculated total DM vector vanishes, indicating that the Dzyaloshinskii--Moriya interaction is negligible compared to the isotropic exchange $J$.

\begin{table}[t!]
    \centering
    \caption{Representative Dzyaloshinskii--Moriya vectors $\mathbf{D}_{mn}$ and their magnitudes (in K) for Na$_6$Cu$_7$BiO$_4$(PO$_4$)$_4$Cl$_3$ at $U=6.66$ eV. One bond is shown for each symmetry-equivalent class, with the corresponding notation indicated in parentheses; all other equivalent DM vectors are generated by the crystallographic symmetry operations of space group $P4/nmm$. The bond labels refer to Fig.~\ref{Na6Cu7_exchanges}. The complete set of DM vectors is given in Tables~\ref{table:D_for_Na6Cu7} and~\ref{table:abs_D_for_Na6Cu7}.}
    \begin{ruledtabular}
    \begin{tabular}{lccr}
    Bond m-n (not.) & \textbf{R$_{mn}$} & \textbf{D$_{mn}$} & $|\textbf{D}|$ \\
    \midrule
    7--11$(\square)$ & ($-2.2$; 2.2; 0.0) & (5.8; 5.8; $-11.2$)  & 13.9 \\
    3--11$(\triangle)$ & ($-2.5$; $-0.3$; 2.1) & (7.3; $-19.1$; $-18.8$)  & 27.7\\
    11--13$(\times)$ & (0.0; $-4.4$; 0.0) & (8.9; 0.0; 0.0) & 8.9 \\
    3--5$(-)$ & ($-5.0$; 0.0; 0.0) & (0.0; $-10.8$; 1.9) & 11.0 \\
    2--11$(+)$ & ($-5.0$;$-2.8$;$-1.8$) & ($-1.7$; $-0.9$; $-2.8$) & 3.3 \\
    \end{tabular}
    \end{ruledtabular}
    \label{table:D_for_Na6Cu7_1bond}
\end{table}

\section{Schwinger-boson mean-field theory and numerical protocol}
\label{sec:sbmft}

Schwinger-boson mean-field theory (SBMFT) provides a unified framework to describe quantum-disordered regimes and magnetically ordered phases in frustrated quantum magnets. In particular, it gives direct access to the excitation spectrum of the quadratic bosonic Hamiltonian and therefore enables the computation of experimentally relevant observables such as equal-time and dynamical spin structure factors. The formalism also naturally accommodates explicit SU(2) symmetry breaking, as induced by Dzyaloshinskii--Moriya (DM) interactions, through the inclusion of triplet hopping and pairing channels. Further methodological details can be found in Refs.~\cite{merino_role_2018,lugan_topological_2019,lugan_schwinger_2022,halimeh_spin_2016,schneider_projective_2022} and references therein.

\subsection{Formalism for Heisenberg + DM interactions}
\label{sec:sbmft_formalism}

In SBMFT the spin operator at site $i$ is represented as
\begin{equation}
\hat{\mathbf{S}}_i = \frac{1}{2}\,\hat{b}_i^\dagger \vec{\sigma}\, \hat{b}_i\,,
\end{equation}
where $\hat{b}_i^\dagger = (\hat{b}_{i\uparrow}^\dagger,\hat{b}_{i\downarrow}^\dagger)$ and $\vec{\sigma}=(\sigma^1,\sigma^2,\sigma^3)$ are the Pauli matrices. Physical spin-$S$ states are recovered by imposing the local constraint
\begin{equation}
\hat{n}_i = \hat{b}_i^\dagger \hat{b}_i = 2S \equiv \kappa\,,
\end{equation}
enforced on average via Lagrange multipliers $\lambda_i$\,.

We consider a Heisenberg--Dzyaloshinskii--Moriya Hamiltonian
\begin{equation}
\mathcal{H}_{\mathrm{HDM}} =
\sum_{i,j} J_{ij}\, \hat{\mathbf{S}}_i \cdot \hat{\mathbf{S}}_j
+ \sum_{i,j} \mathbf{D}_{ij} \cdot
\left( \hat{\mathbf{S}}_i \times \hat{\mathbf{S}}_j \right).
\label{eq:H_HDM}
\end{equation}
To treat SU(2)-symmetric and SU(2)-breaking terms on equal footing, we introduce singlet and triplet hopping/pairing operators
\begin{align}
\hat{h}_{ij}^\alpha &= \frac{1}{2}\, \hat{b}_i^\dagger \sigma^\alpha \hat{b}_j, \\
\hat{p}_{ij}^\alpha &= \frac{i}{2}\, \hat{b}_i\left( \sigma^\alpha \sigma^2 \right) \hat{b}_j\,,
\end{align}
with $\alpha=0$ the singlet channel ($\sigma^0=\mathbb{I}$) and $\alpha=1,2,3$ the triplet channels. Using vector notations
${\hat{\mathbf{h}}}_{ij} = (\hat{h}_{ij}^0,\hat{h}_{ij}^1,\hat{h}_{ij}^2,\hat{h}_{ij}^3)^T$
and
${\hat{\mathbf{p}}}_{ij} = (\hat{p}_{ij}^0,\hat{p}_{ij}^1,\hat{p}_{ij}^2,\hat{p}_{ij}^3)^T$,
the exchange terms can be rewritten as
\begin{align}
\hat{\mathbf{S}}_i \cdot \hat{\mathbf{S}}_j
&= :{\hat{\mathbf{h}}}_{ij}^\dagger m_0 {\hat{\mathbf{h}}}_{ij}:
- {\hat{\mathbf{p}}}_{ij}^\dagger m_0 {\hat{\mathbf{p}}}_{ij}, \\
\mathbf{D}_{ij} \cdot
\left( \hat{\mathbf{S}}_i \times \hat{\mathbf{S}}_j \right)
&=
:{\hat{\mathbf{h}}}_{ij}^\dagger m_{ij}^{\mathrm{DM}} {\hat{\mathbf{h}}}_{ij}:
- {\hat{\mathbf{p}}}_{ij}^\dagger m_{ij}^{\mathrm{DM}} {\hat{\mathbf{p}}}_{ij}\,.
\end{align}
Here $m_0=\mathrm{diag}(1,-\tfrac{1}{3},-\tfrac{1}{3},-\tfrac{1}{3})$, while $m_{ij}^{\mathrm{DM}}$ is an antisymmetric $4\times4$ matrix encoding the orientation and magnitude of $\mathbf{D}_{ij}$ (as defined in Sec.~\ref{dm}). The DM interaction thus couples naturally to triplet channels and explicitly breaks SU(2) symmetry.

The resulting Schwinger-boson Hamiltonian reads
\begin{align}
\mathcal{H}_{\mathrm{SB}}
&=
\sum_{i,j}
\Big[
:{\hat{\mathbf{h}}}_{ij}^\dagger H_{ij} {\hat{\mathbf{h}}}_{ij}:
+ {\hat{\mathbf{p}}}_{ij}^\dagger P_{ij} {\hat{\mathbf{p}}}_{ij}
\Big]
+ \sum_i \lambda_i (\hat{n}_i - \kappa)\,,
\label{eq:H_SB}
\end{align}
with $H_{ij}=J_{ij} m_0 + m_{ij}^{\mathrm{DM}}$ and $P_{ij}=-H_{ij}$. A mean-field decoupling is performed by introducing $\mathbf{h}_{ij}=\langle \hat{\mathbf{h}}_{ij}\rangle$ and $\mathbf{p}_{ij}=\langle \hat{\mathbf{p}}_{ij}\rangle$, evaluated in the bosonic vacuum $|\phi_0\rangle$. The quadratic Hamiltonian is diagonalized via a bosonic Bogoliubov transformation, yielding the spinon bands and enabling the computation of correlation functions.

\subsection{Numerical protocol, stability checks, and observables}
\label{sec:sbmft_numerics}

\begin{figure}
    \centering
    \includegraphics[width=0.6\linewidth]{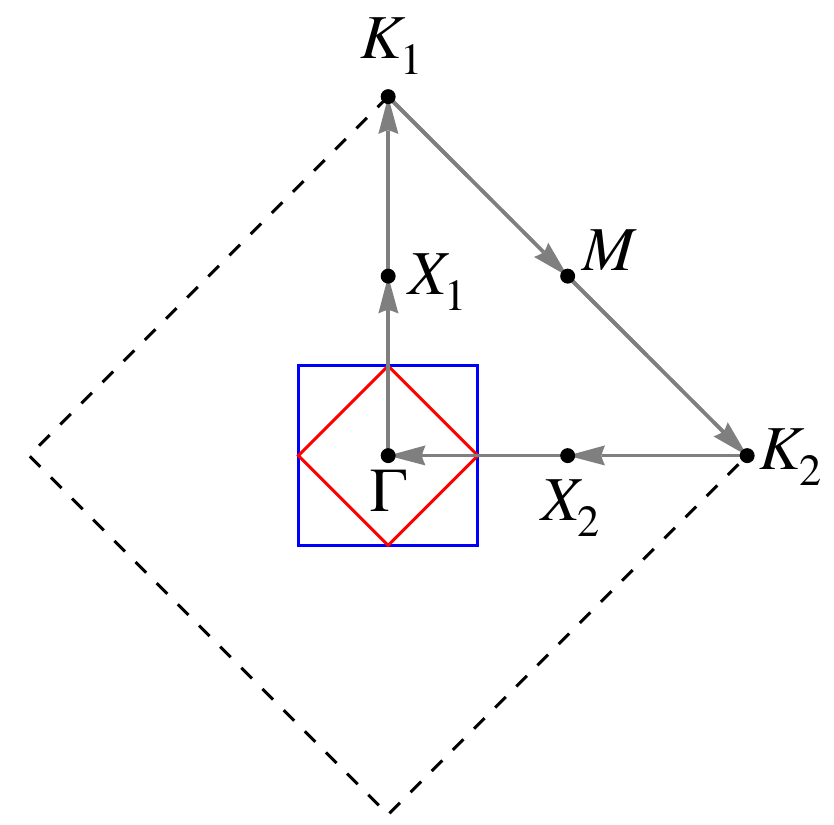}
\caption{First Brillouin zones corresponding to the different unit cells considered in this work: the square--kagome lattice with $n_u=6$ sites per unit cell (blue), the enlarged unit cell with $n_u=14$ sites (red), and the underlying square lattice (black dashed). High-symmetry points are indicated, and the gray arrows denote the momentum path used to compute the dynamical spin structure factors.}
    \label{sbmft0}
\end{figure}

Self-consistency was initialized from multiple starting configurations, including (i) fully random mean-field amplitudes and (ii) symmetry-distinct flux seeds used in the square--kagome SBMFT literature. The converged solutions reported in this work correspond to stable fixed points of the self-consistency cycle: small perturbations of the converged amplitudes systematically flow back to the same solution upon iteration. Calculations were performed on clusters of size $N=n_u \times l^2$ sites, with $n_u$ the number of sites in the unit-cell, up to $N=1944$ spins for the isotropic square--kagome model ($n_u=6$, $l=18$) and up to $N=4032$ sites for the realistic compound Hamiltonian ($n_u=28$, $l=12$) in the regime of small $J_{10}$. The momentum resolution and spectral broadening used for the dynamical structure factors were checked such that the qualitative features discussed below are insensitive to these numerical parameters. The Brillouin zones corresponding to the two unit cells, together with the high-symmetry points and the momentum paths used to compute the dynamical structure factors, are shown in Fig.~\ref{sbmft0}.

We work on finite clusters, which introduce a small finite-size spinon gap even when the thermodynamic system is expected to condense. This controlled finite-size gap prevents immediate Bose condensation up to a certain critical $S$ and allows us to monitor ordering tendencies continuously through clear precursors of mode softening, either via the suppression of the minimum spinon gap or through the redistribution of spectral weight in the dynamical structure factor.

To characterize gauge-inequivalent mean-field states, we use Wilson-loop (WL) fluxes constructed from the complex pairing amplitudes \cite{wang_spin-liquid_2006,tchernyshyov_flux_2006}. On the square--kagome lattice a natural choice is to consider elementary six-site loops \cite{lugan_topological_2019}. For a generic loop $\mathcal{C}=(i\!\to\! j\!\to\! k\!\to\! l\!\to\! m\!\to\! n\!\to\! i)$, the associated WL is defined as
\begin{equation}
W_{\mathcal{C}} =
\arg\!\left[
p_{ij}^0\,
(-p^{0*}_{jk})\,
p_{kl}^0\,
(-p^{0*}_{lm})\,
p_{mn}^0\,
(-p^{0*}_{ni})
\right],
\end{equation}
which is invariant under local gauge transformations and therefore provides a gauge-independent flux threading the loop. Each SBMFT saddle point can then be labeled by the set of inequivalent WL fluxes within a unit cell, which we denote as $\Phi = (\phi_1,\phi_2,\phi_3,\phi_4)$ where each $\phi_\mu$ is the WL flux modulo $2\pi$.

The results of applying this framework to (i) the isotropic square--kagome model, (ii) the minimal DM perturbation test, and (iii) the compound-specific Hamiltonian (including the $J_{10}$-tuning analysis and the impact of symmetry-allowed DM vectors) are presented in Sec.~\ref{results}.

\begin{figure}
    \centering
    \includegraphics[width=0.9\linewidth]{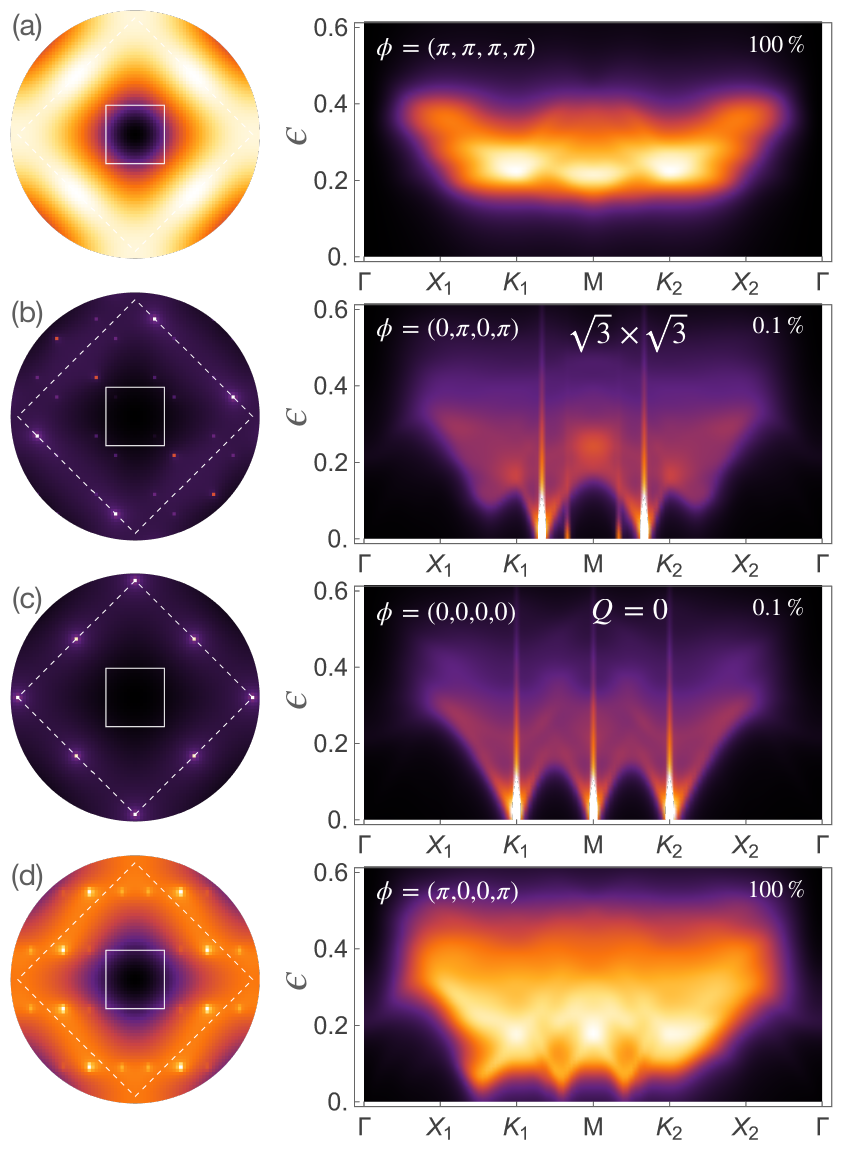}
\caption{Equal-time (left) and dynamical (right) spin structure factors obtained within SBMFT for the four lowest-energy Ans\"atze of the isotropic square-kagome model at $S=0.3$ without Dzyaloshinskii--Moriya interactions. Panels (a)--(d) correspond to the Wilson-loop flux patterns $\Phi=(\pi,\pi,\pi,\pi)$, $(0,\pi,0,\pi)$, $(0,0,0,0)$, and $(\pi,0,0,\pi)$, identified respectively as a gapped quantum spin liquid, the $\sqrt{3}\!\times\!\sqrt{3}$ order, the $Q=0$ order, and a second gapped quantum spin liquid. The percentage of the total spectral weight displayed is indicated in the plots. The first (solid line) and extended (dashed line) Brillouin zones of the square-kagome lattice showing the location of the Bragg peaks with the fraction of the total spectral weight of the classical (a) $\mathbf{Q}=0$ order, at $(4\pi,0)$ and $(2\pi,2\pi)$ [and symmetry related points] with equal spectral weight, and (b) $\sqrt{3}\times\sqrt{3}$ order, at $(2\pi \pm q, 2\pi \mp q)$ with $q = 4\pi/3$, and leading subdominant peaks at $(q,-q)$ with 36\%, i.\,e., $\lambda/\Lambda=0.36$, of the spectral weight of the dominant ones~\cite{note_peaks}. The $\sqrt{3}\times\sqrt{3}$ order breaks the four-fold rotational symmetry.}
    \label{sbmft1}
\end{figure}

\begin{figure}
    \centering
    \includegraphics[width=0.65\linewidth]{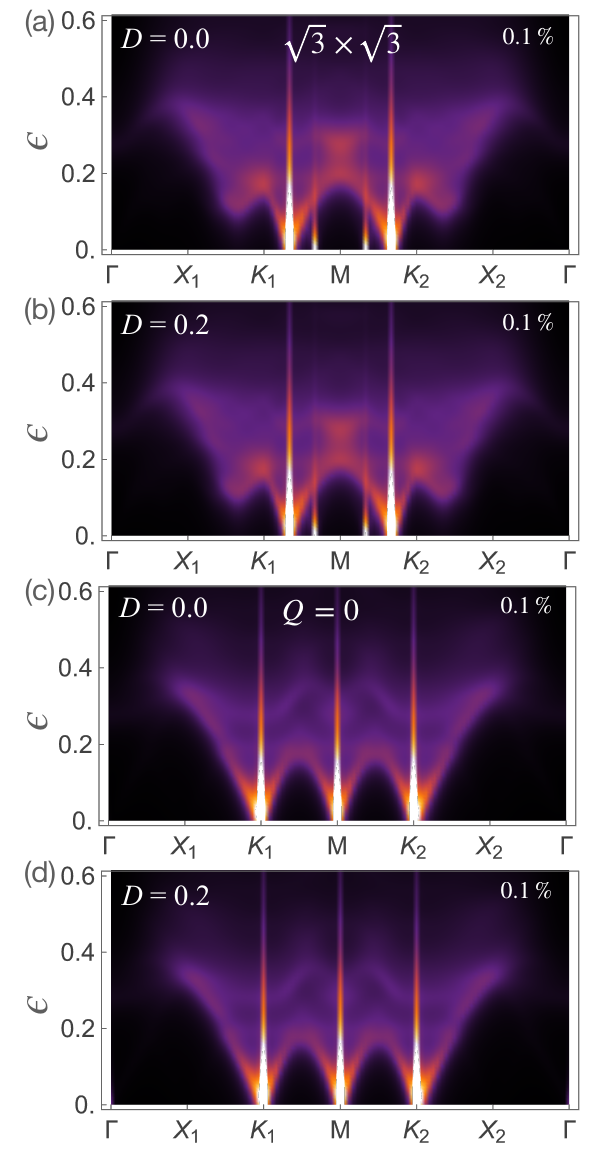}
\caption{Dynamical spin structure factor of the isotropic square-kagome model at $S=(\sqrt{3}-1)/2\simeq0.366$. Panels (a,b) show the $\sqrt{3}\!\times\!\sqrt{3}$ state and (c,d) the $Q=0$ state, without and with DM interactions on the $J_2$ bonds, respectively. The percentage of the total spectral weight displayed is indicated in the plots.}
    \label{structfactcompisodm}
\end{figure}

\section{Results of SBMFT calculations}
\label{results}

\subsection{Isotropic square--kagome model: competing SBMFT saddle points and their signatures}

We first benchmark our generalized Schwinger-boson mean-field framework on the \emph{isotropic} square--kagome Heisenberg antiferromagnet, where the theoretical expectation is the existence of several nearby competing phases (ordered and quantum-disordered) within the same lattice geometry~\cite{ralko_resonating-valence-bond_2015,Jahromi2025,Richter2022,Niggemann2023,lugan_topological_2019,Richter2023,Morita2018,Derzhko2014}. We solve the self-consistent SBMFT equations by tuning the effective spin parameter $\kappa=2S$. The physical value for the square--kagome lattice corresponds to $S(\sqrt{3}-1)/2\simeq0.366$, which restores the equal-time structure-factor sum rule within SBMFT \cite{lugan_topological_2019}. As commonly done in SBMFT studies of frustrated magnets, we slightly reduce the effective spin in order to enhance quantum fluctuations and stabilize gapped mean-field solutions, allowing the parent quantum spin-liquid Ans\"atze to be identified unambiguously. In the following we therefore use $\kappa=0.6$ (i.e., $S=0.3$).

At this value, four distinct low-energy self-consistent solutions are obtained, which can be unambiguously classified by their gauge-invariant Wilson-loop (WL) flux patterns on elementary loops \cite{wang_spin-liquid_2006,tchernyshyov_flux_2006,lugan_topological_2019}. The WLs discussed in the previous section are defined on elementary six-site loops $(4\to13\to6\to7\to11\to9\to4)$ of  Fig.~\ref{Na6Cu7_exchanges}. Each unit cell admits four symmetry-inequivalent orientations of such loops. The four competing saddle points are characterized by
\begin{equation}
\Phi=(\pi,\pi,\pi,\pi),(0,\pi,0,\pi),(0,0,0,0),(\pi,0,0,\pi)\,.
\end{equation}

The corresponding equal-time and dynamical spin structure factors are shown in Fig.~\ref{sbmft1}, providing experimentally relevant fingerprints of each saddle point. For clarity, the spectra are normalized to their total weight; in strongly ordered regimes, where most of the weight condenses at the ordering wave vector, only a fraction of the remaining spectral weight is displayed, as indicated in the plots. This representation highlights the strength of the order, reveals residual spectral features, and facilitates comparison between different system sizes.
A key outcome of Fig.~\ref{sbmft1} is that the four saddle points split into two qualitatively distinct classes:
(i) two quantum-disordered solutions with fully gapped spinon spectra and diffuse momentum-space correlations [Fig.~\ref{sbmft1}(a,d)], and
(ii) two magnetically ordered solutions signaled, within SBMFT, by pronounced low-energy spectral weight at the ordering wavevectors expected for the $Q=0$ states and $\sqrt{3}\!\times\!\sqrt{3}$ states~\cite{Richter-2009} and sharp features in the equal-time structure factor~\cite{Astrakhantsev-2021} [Fig.~\ref{sbmft1}(b,c)]. In the present parameter set, the magnetically ordered solutions are energetically favored: the $\sqrt{3}\!\times\!\sqrt{3}$ and $Q=0$ states lie below the gapped saddle points, with the $\sqrt{3}\!\times\!\sqrt{3}$ state being the lowest-energy solution in our unrestricted search.

Throughout this work, energies are reported as dimensionless mean-field energies per site, expressed in units of the largest exchange coupling $J_2$. For the four Ans\"atze discussed above, we obtain
$
\epsilon_{(\pi,\pi,\pi,\pi)}=-0.2251(1),
\epsilon_{(0,\pi,0,\pi)}=-0.2315(1),
\epsilon_{(0,0,0,0)}=-0.2313(6),
\epsilon_{(\pi,0,0,\pi)}=-0.2293(5).
$
The two ordered solutions $(0,\pi,0,\pi)$ and $(0,0,0,0)$ are nearly degenerate at this value of $\kappa$, which motivates a controlled study of how weak SU(2)-breaking perturbations, such as Dzyaloshinskii--Moriya interactions, affect the ordering tendencies. The same energy convention is used consistently throughout the paper, including in the analysis of the dynamical structure factors.

\subsection{Minimal DM perturbation test: DM on $J_2$ bonds preserves the qualitative hierarchy}

To isolate the \emph{intrinsic} role of DM interactions in the square--kagome geometry before turning to the compound-specific Hamiltonian, we perform a minimal diagnostic in which DM terms are applied only on the $J_2$ bonds of the isotropic model. This choice is not meant to represent a specific material; rather, it is designed as a clean perturbation that (i) explicitly breaks SU(2), (ii) activates triplet hopping/pairing channels in the generalized SBMFT decoupling, and (iii) tests the stability of the competing low-energy saddle points against anisotropy.

For this minimal DM test, we now work at the effective spin parameter $S=(\sqrt{3}-1)/2\simeq 0.366$ mentioned earlier. Figure~\ref{structfactcompisodm} shows the dynamical structure factor for the two low-energy ordered solutions ($\sqrt{3}\!\times\!\sqrt{3}$ and $Q=0$) without DM ($D/J=0$) and with a representative anisotropy ($D/J=0.2$). Two robust observations follow: (i)) Introducing DM on the $J_2$ bonds does \emph{not} generate any new competing phase within the explored window: the momentum-space structure of correlations remains qualitatively unchanged. (ii) DM slightly strengthens ordering tendencies (as visible in the enhanced low-energy spectral weight at the same ordering wavevectors), and the energetic splitting remains small but the hierarchy is preserved: for $D/J=0.2$ the dimensionless energies per site shift from  $\epsilon_{\sqrt{3}\times\sqrt{3}}=-0.3104(5))$ and $\epsilon_{Q=0}=-0.3102(4)$ to $\epsilon_{\sqrt{3}\times\sqrt{3}}=-0.3135(2)$ and $\epsilon_{Q=0}=-0.3133(3)$, confirming that weak-to-moderate DM anisotropy does not qualitatively reorder the competition in this minimal setting. This \emph{robustness check} is important because it establishes that the dramatic DM effects discussed below for the realistic Hamiltonian arise from the combined influence of lattice decoration, exchange hierarchy (notably $J_{10}$), and the full symmetry-allowed DM pattern, rather than from a fragile instability of the isotropic square--kagome SBMFT solutions.

\subsection{Realistic Hamiltonian for Na$_6$Cu$_7$BiO$_4$(PO$_4$)$_4$Cl$_3$: DM vectors and the central role of $J_{10}$}

We now turn to the compound-specific Hamiltonian for Na$_6$Cu$_7$BiO$_4$(PO$_4$)$_4$Cl$_3$, where the square--kagome network formed by Cu(1) and Cu(2) is decorated by Cu(3) sites in a checkerboard fashion. From the DFT-based analysis, the dominant isotropic exchanges are $J_1$ and $J_2$ on the square--kagome backbone, with additional frustrating diagonal and longer-range couplings (including $J_3$ and $J_4$) and, crucially, a sizeable coupling $J_{10}$ connecting the decorating Cu(3) sites to the backbone (Table~\ref{table:J_for_Na6Cu7_1} and Fig.~\ref{Na6Cu7_exchanges}). The $J_{10}$ coupling controls the degree of hybridization between the decorating sublattice and the frustrated backbone. The symmetry-allowed DM vectors are extracted from spin-orbit-coupled Wannier-based hopping matrices and summarized (one representative per symmetry class) in Table~\ref{table:D_for_Na6Cu7}. Their magnitudes reach $\sim 10$--$30$ K on the leading exchange pathways, making them a potentially relevant perturbation on the energy scales of interest.

\begin{figure}[h!]
    \centering
    \includegraphics[width=0.9\textwidth]{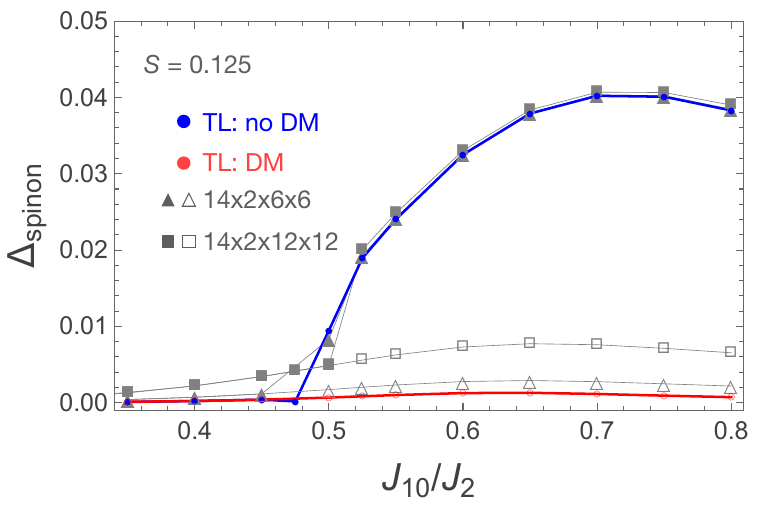}
    \caption{Minimum spinon gap $\Delta_{\mathrm{spinon}}$ as a function of $J_{10}/J_2$ for the realistic exchange parameters of Ref.~\cite{Niggemann2023} at $S=0.125$. Open and filled symbols denote calculations without and with DM interactions, respectively, for finite clusters of size $N=1008$ (triangles) and $N=4032$ (squares). Blue and red circles indicate the corresponding thermodynamic-limit estimates obtained from finite-size scaling. Reducing $J_{10}$ suppresses the spinon gap, while DM interactions shift the system further toward condensation.}
    \label{evolspinongap}
\end{figure}

Within SBMFT, $J_{10}$ emerges as an efficient control parameter for the proximity to boson condensation (magnetic order). To quantify this statement we compute the minimum spinon gap $\Delta_{\rm spinon}$ as a function of $J_{10}/J_2$ for the realistic exchange parameters of Ref.~\cite{Niggemann2023} at fixed $S=0.125$ (chosen to stabilize gapped solutions and enable controlled tracking of softening trends). Fig.~\ref{evolspinongap} shows $\Delta_{\rm spinon}(J_{10})$
for two cluster sizes $N=1008$ and $4032$ (respectively $l=6$ and $12$), together with a thermodynamic-limit extrapolation obtained from finite-size scaling.

Three strong conclusions follow directly from Fig.~\ref{evolspinongap}:

(i) $J_{10}$ stabilizes the gapped regime.
Reducing $J_{10}$ suppresses $\Delta_{\rm spinon}$ continuously, driving the system toward a condensation threshold within SBMFT.

(ii) DM promotes condensation.
Including the full symmetry-allowed DM pattern shifts the gap downward relative to the $D=0$ case for the same $J_{10}$, indicating that spin--orbit-induced anisotropy enhances the tendency toward magnetic ordering in the realistic model.

(iii) The trend is not a finite-size artifact.
The finite-size data at $N=1008$ and $N=4032$ exhibit a consistent systematic evolution, and the extrapolated thermodynamic-limit curves retain the same qualitative separation between the $D=0$ and $D\neq0$ cases, supporting the robustness of the conclusion that $J_{10}$ tunes the system through a near-critical regime and that DM shifts the instability toward larger $J_{10}$.

\subsection{Spectral fingerprints: illustrative vs realistic $J_{10}$ and the effect of DM}

\begin{figure}[ht]
    \centering
    \includegraphics[width=0.9\textwidth]{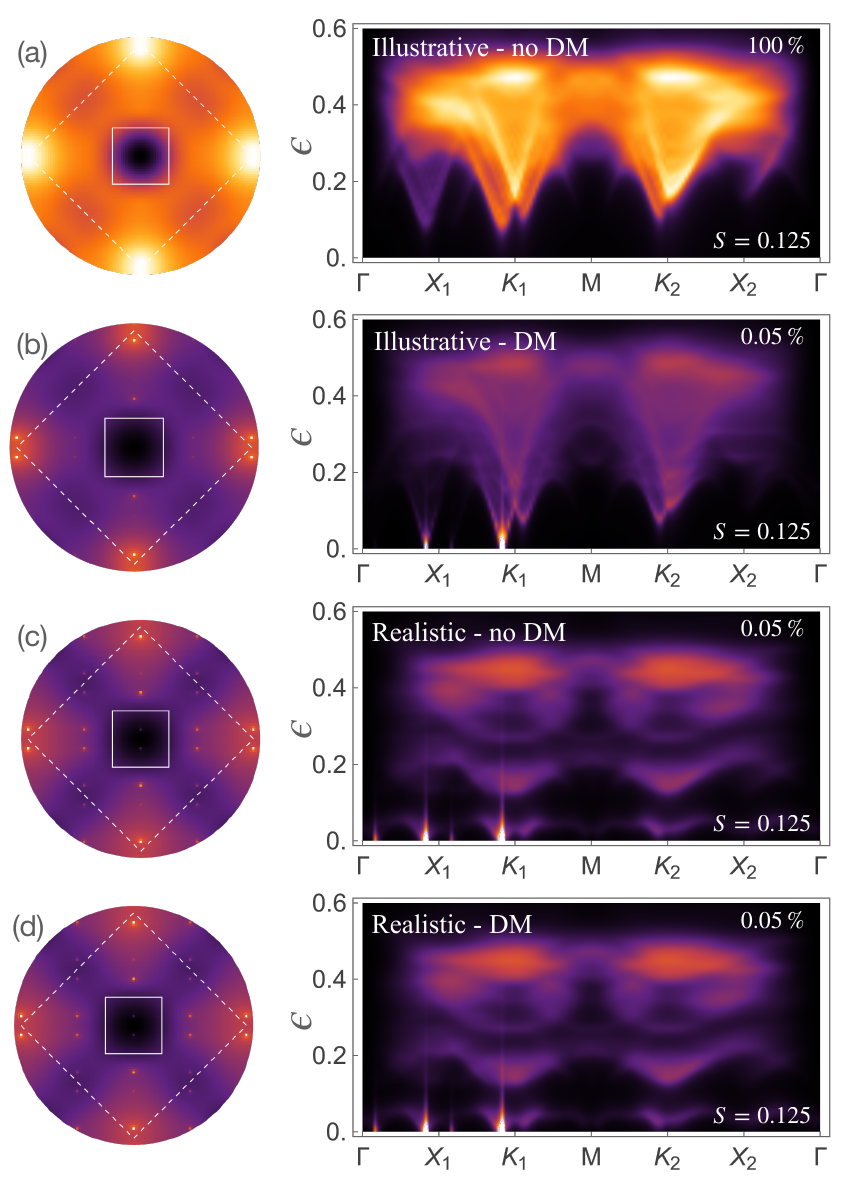}
\caption{Equal-time (left) and dynamical (right) spin structure factors for the DFT-derived exchange parameters of Ref.~\cite{Niggemann2023} at $S=0.125$. Panels (a,b) show an illustrative gapped regime with $J_{10}=150$ K, without and with DM interactions, respectively; panels (c,d) show the realistic regime with $J_{10}=64.6$ K as calculated in Ref.~\cite{Niggemann2023}), again without and with DM. DM vectors are taken from Table~\ref{table:D_for_Na6Cu7}. Lowering $J_{10}$ transfers spectral weight to low energies, a tendency further enhanced by the symmetry-allowed DM interactions. The percentage of the total spectral weight shown in each panel is indicated in the plots.}
    \label{oldjstruct}
\end{figure}

To connect the gap evolution to experimentally relevant correlation functions, we compute both equal-time and dynamical spin structure factors for two representative regimes (Fig.~\ref{oldjstruct}).

The illustrative regime uses an artificially large $J_{10}=150$ K and therefore provides a controlled reference deep in the gapped, quantum-disordered side of the SBMFT solutions. In this regime, both the equal-time maps and the dynamical response exhibit a clear finite-energy onset and broad momentum-space features consistent with a stable gapped spectrum.

The realistic regime uses the compound parameter $J_{10}=64.6$ K (as in Ref.~\cite{Niggemann2023}) and reveals the main qualitative change: spectral weight is transferred toward low energies at specific momenta, signaling pronounced soft-mode formation. Importantly, adding the symmetry-allowed DM vectors (taken from Table~\ref{table:D_for_Na6Cu7}) further enhances this low-energy transfer, consistent with the DM-driven downward shift of $\Delta_{\rm spinon}$ seen in Fig.~\ref{evolspinongap}. Taken together, Figs.~\ref{evolspinongap}--\ref{oldjstruct} establish a coherent picture: $J_{10}$ controls the stability of the gapped regime, and DM interactions enhance the instability toward condensation in the realistic decorated square--kagome Hamiltonian.

A noteworthy aspect of the realistic spectra is that none of the four flux states stabilized in the isotropic square--kagome benchmark appears to describe the compound Hamiltonian in a simple way. Instead, the location of the soft modes and the redistribution of spectral weight point to a more complex ordering tendency in the decorated model, consistent with the additional Cu(3) sublattice and the dominant exchange scales.

\subsection{Alternative exchange set and proximity to SBMFT condensation}

\begin{figure}
    \centering
    \includegraphics[width=0.9\textwidth]{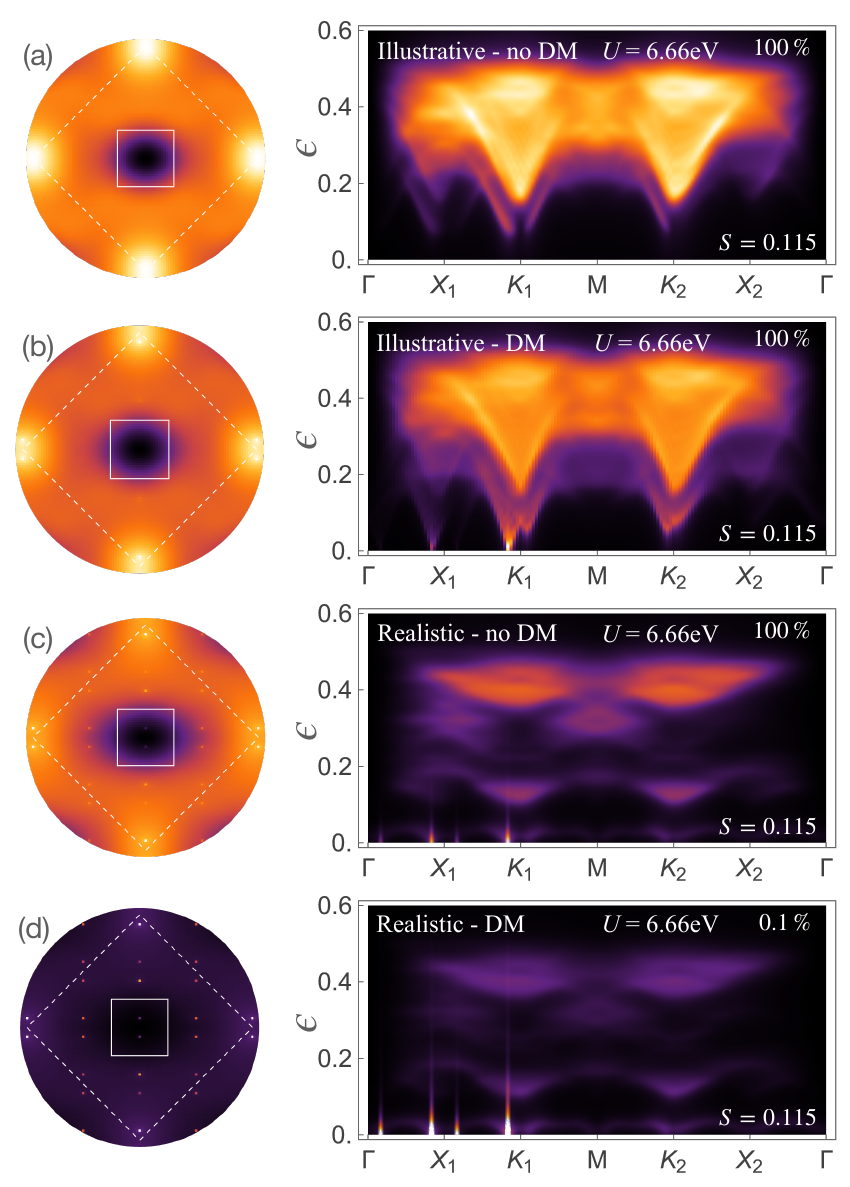}
\caption{Equal-time (left) and dynamical (right) spin structure factors for the exchange parameters obtained at $U=6.66$ eV, evaluated at $S=0.115$. Panels (a,b) show an illustrative regime with an arbitrarily large $J_{10}$, without and with DM interactions, respectively; panels (c,d) show the realistic regime with $J_{10}=64.6$ K, again without and with DM. DM vectors are taken from Table~\ref{table:D_for_Na6Cu7}. The enhanced low-energy spectral weight indicates that this parameter set lies closer to magnetic condensation. The percentage of the total spectral weight displayed in each panel is indicated in the plots.}
    \label{newjstruct}
\end{figure}

Finally, we repeat the analysis for the alternative exchange parameters obtained for $U=6.66$ eV (Table~\ref{table:J_for_Na6Cu7_1}), using $S=0.115$ and comparing illustrative and realistic regimes with and without DM and on a cluster of $N=4032$ sites (Fig.~\ref{newjstruct}). In this parameter set, $J_{10}$ is smaller, placing the system even closer to the condensation boundary. As a result, including DM interactions makes self-consistency increasingly demanding near the physical $J_{10}$: the iterative cycle tends to drive the solution toward condensation, and convergence can fail in the immediate vicinity of the physical parameter set. Nevertheless, convergence is achieved for slightly larger $J_{10}$, and the \emph{smooth evolution} of the converged solutions toward the physical regime supports a physically controlled interpretation: the lack of convergence is not evidence of an uncontrolled numerical artifact, but rather is consistent with the expected breakdown of a gapped SBMFT saddle point as the system approaches boson condensation (magnetic ordering) in the thermodynamic limit.

\section{Discussion and outlook}
\label{conclusion}

We have investigated the role of Dzyaloshinskii--Moriya (DM) interactions in square-kagome quantum antiferromagnets by combining \emph{ab initio} extraction of symmetry-allowed DM vectors for Na$_6$Cu$_7$BiO$_4$(PO$_4$)$_4$Cl$_3$ with a generalized Schwinger-boson self-consistent mean-field treatment that retains both singlet and triplet hopping/pairing channels. This combination allows us to address, within a single framework, both the generic effect of anisotropy on the square-kagome geometry and its material-specific impact in a decorated square-kagome compound.

At the level of the isotropic square-kagome benchmark, we identified several competing low-energy SBMFT saddle points distinguished by their Wilson-loop fluxes and by clear equal-time and dynamical structure-factor fingerprints. A minimal DM perturbation restricted to the $J_2$ bonds does not qualitatively reshape this competing landscape, but it already enhances ordering tendencies, indicating that anisotropy acts in a systematic rather than accidental way in this geometry.

For the realistic compound Hamiltonian of Na$_6$Cu$_7$BiO$_4$(PO$_4$)$_4$Cl$_3$, a clearer physical picture can be established. The coupling $J_{10}$, which links the decorating Cu(3) sites to the square-kagome backbone, acts as the control parameter stabilizing the gapped quantum-paramagnetic regime. Reducing $J_{10}$ continuously suppresses the minimum spinon gap, while the full symmetry-allowed DM pattern shifts the system further toward boson condensation. In this sense, decorated square-kagome antiferromagnets exhibit a simple but important control knob: $J_{10}$ stabilizes the quantum-paramagnetic regime, whereas symmetry-allowed DM interactions systematically destabilize it and drive the system toward magnetic condensation.

This places Na$_6$Cu$_7$BiO$_4$(PO$_4$)$_4$Cl$_3$ in close proximity to a magnetic instability. Our results therefore refine the picture of this compound from that of a merely robust quantum paramagnet to that of a material located near the boundary between a gapped disordered regime and anisotropy-enhanced ordering. More broadly, our results show that decorated square-kagome materials are not simply Heisenberg-frustrated magnets, but systems in which spin--orbit anisotropy provides a quantitatively relevant tuning field.

The present work also suggests several experimental directions. Since the leading DM terms enhance low-energy soft modes and redistribute spectral weight in momentum space, anisotropy-sensitive probes such as NMR and direction-dependent susceptibility measurements should be particularly informative. ESR may also be useful provided that the relevant low-energy anisotropy scale lies within the experimentally accessible frequency window. If momentum-resolved spectroscopy becomes available, the predicted buildup of low-energy spectral weight at specific wave vectors would provide a direct test of the proximity to condensation identified here. On the theory side, it will be important to complement the present SBMFT analysis with approaches that treat the onset of order beyond mean field and can further clarify the nature of the instability in the immediate vicinity of the physical parameter regime.

Overall, our results establish DM interactions as a central ingredient of square-kagome quantum magnetism in realistic low-symmetry materials. In doing so, they identify a general instability mechanism for decorated square-kagome antiferromagnets and provide a microscopic framework for interpreting the growing family of square-kagome and related materials.

\section*{Acknowledgements}
We thank Harald O. Jeschke and Alexander N. Vasiliev for helpful discussions. 

The work Y.I. was performed in part at the Aspen Center for Physics, which is supported by a grant from the Simons Foundation (1161654, Troyer). This research was also supported in part by grant NSF PHY-2309135 to the Kavli Institute for Theoretical Physics and by the International Centre for Theoretical Sciences (ICTS) for participating in the Discussion Meeting - Fractionalized Quantum Matter (code: ICTS/DMFQM2025/07). Y.I. acknowledges support from the Abdus Salam International Centre for Theoretical Physics through the Associates Programme, from the Simons Foundation through Grant No.~284558FY19, from IIT Madras through the Institute of Eminence program for establishing QuCenDiEM (Project No. SP22231244CPETWOQCDHOC). S.S. and A.R. thank IIT Madras for a Visiting Faculty Fellow position under the IoE program during which this project was initiated. A.R. acknowledges support from the french national agency (ANR FlatMoi, grant no. ANR-21-CE30-0029). L.T. and S.S. acknowledge support of the Ministry of Science and Higher Education of the Russian Federation through the IMP UB RAS. The work of VVM was supported by the Ministry of Science and Higher Education of the Russian Federation (Grant No. FEUZ-2026-0010).


\appendix

\section{Electronic structure}
\label{appendix_A}

The electronic structure obtained within the GGA calculations is presented in the form of the total and orbital-resolved density of states of Na$_6$Cu$_7$BiO$_4$(PO$_4$)$_4$[Cl]$_3$, complementing the band-structure evidence for the isolated Cu $d_{x^2-y^2}$-derived manifold discussed in Sec.~\ref{dm}. The energy range from $-7.8$~eV to $-3.6$~eV below the Fermi level is predominantly occupied by O-$p$ states hybridized with Cu-$d$, P-$p$, and Cl-$p$ orbitals. From $-3.6$~eV to $0.7$~eV, Cu-$d$ states dominate, hybridizing with O-$p$ and Cl-$p$ states. An energy gap is observed at approximately from $-1$ to $-0.6$~eV relative to $E_F$. The strongest peaks in the partial Cu-$d$ DOS appear at $-2.0$~eV and $-1.3$~eV, corresponding to fully filled $xz/yz$, $3z^2 - r^2$, and $xy$ orbitals, as well as the half-filled $x^2 - y^2$ orbital. The unoccupied $x^2 - y^2$ states are located between $-0.6$~eV and $0.6$~eV.

\begin{figure}[h!]
\centering
\includegraphics[width=1\columnwidth]{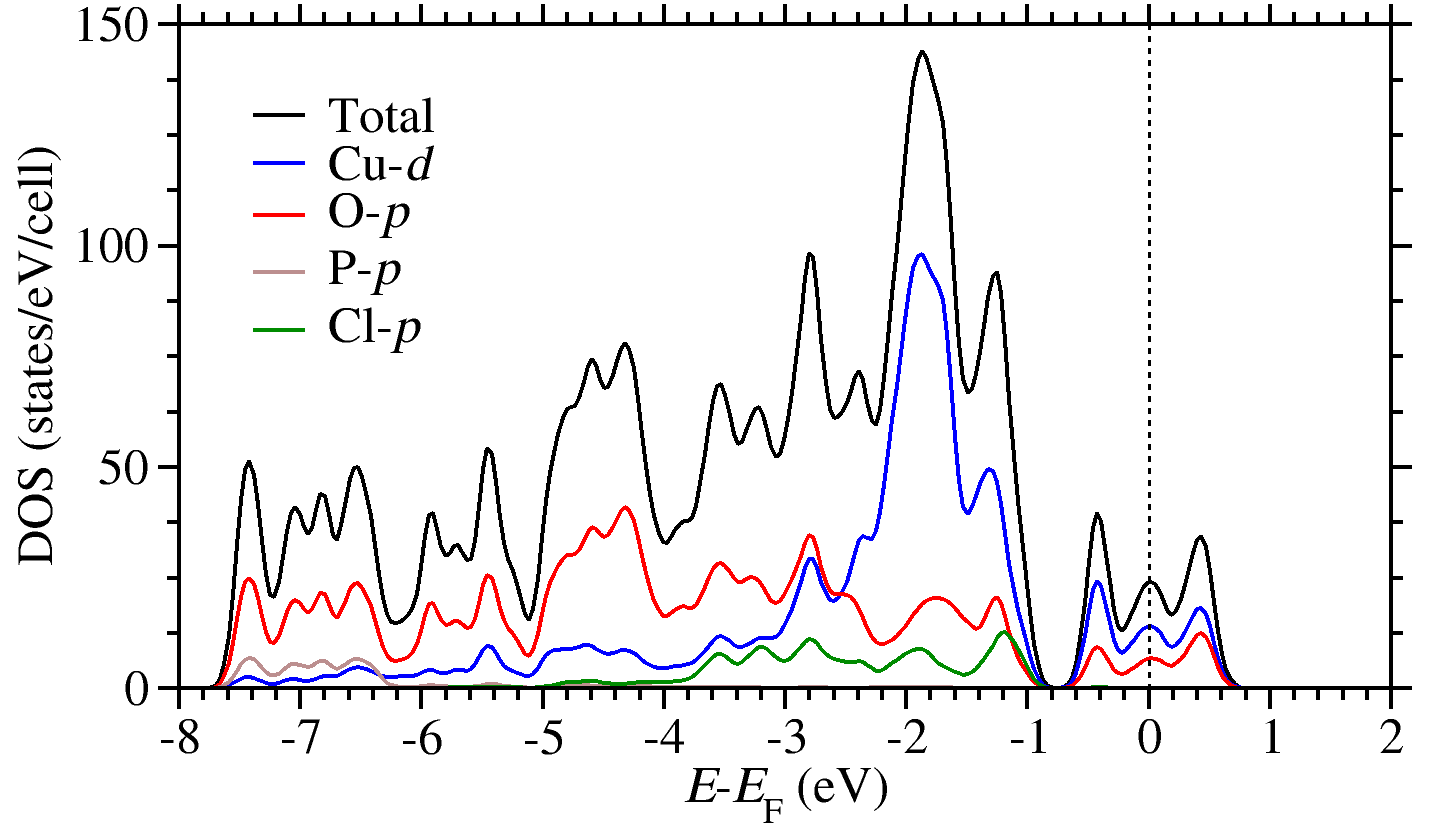}
\caption{Total and orbital-resolved density of states of Na$_6$Cu$_7$BiO$_4$(PO$_4$)$_4$Cl$_3$ from nonmagnetic GGA calculations. The black curve shows the total density of states, while the colored curves show the partial contributions from Cu-$d$, O-$p$, P-$p$, and Cl-$p$ states. The Fermi energy is set to zero. The low-energy unoccupied manifold is dominated by Cu $d_{x^2-y^2}$ character, consistent with the employed one-band description.}
    \label{Na6Cu7_GGA_DOS}
\end{figure}

\newpage

\section{Isotropic and antisymmetric anisotropic exchange interactions}
\label{appendix_B}

\begin{figure}[h!]
\centering
\includegraphics[width=0.9\columnwidth]{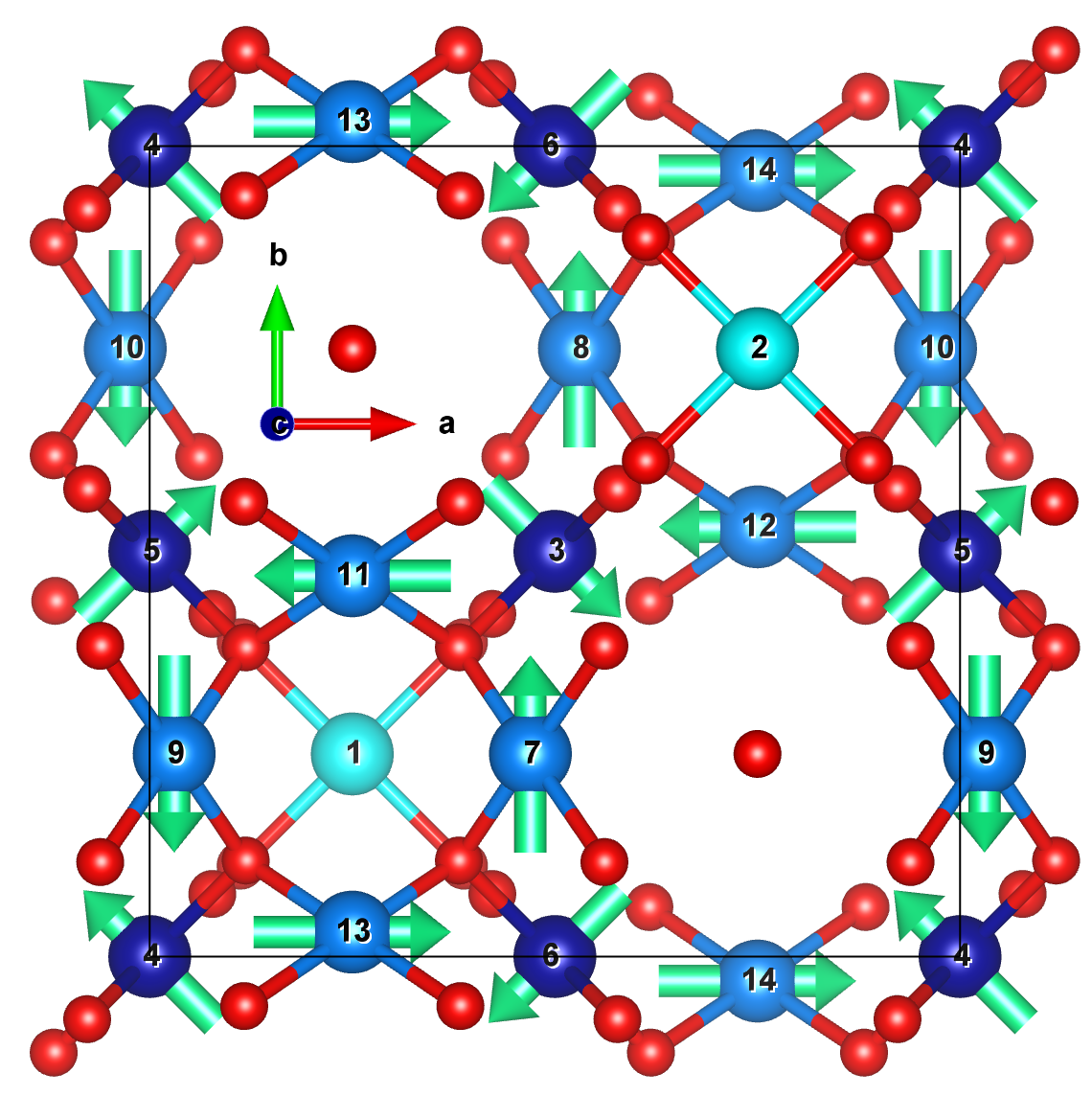}
\caption{Total DM interaction vectors (green arrows) for all Cu sites in Na$_6$Cu$_7$BiO$_4$(PO$_4$)$_4$Cl$_3$. The arrows indicate the site-resolved sums of the bond DM vectors connected to each Cu site, shown here in the crystallographic geometry of the unit cell. The resulting pattern summarizes the local anisotropic environment that constrains the symmetry-allowed canting tendencies of the Cu moments.}
    \label{fig:sum_D_ab}
\end{figure}

\begin{figure}[h!]
\centering
\includegraphics[width=0.88\columnwidth]{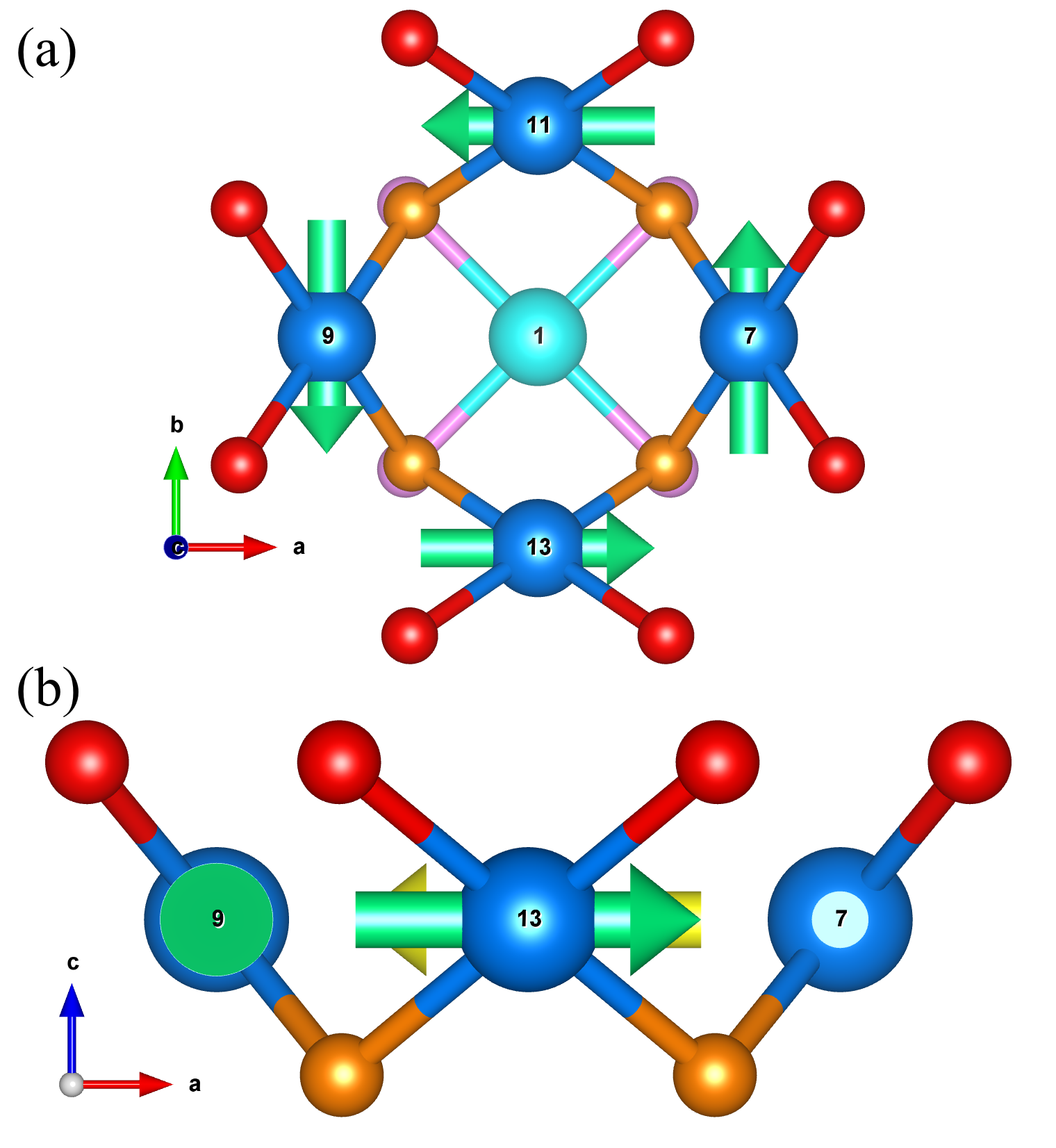}
\caption{Total DM interaction vectors (green arrows) for Cu(2) sites in Na$_6$Cu$_7$BiO$_4$(PO$_4$)$_4$Cl$_3$ viewed along the (a) $c$-axis ($ab$-plane) and (b) $b$-axis ($ac$-plane). Red, orange, and pink spheres denote oxygen atoms at distinct heights along the $c$-axis.}
    \label{fig:sum_D_Cu(2)}
\end{figure}

\onecolumngrid
\begin{longtable}{@{\extracolsep{\fill}} c c c c c c c c c c c @{}}
\caption{Complete set of isotropic exchange interactions $J_n$ (in K) and corresponding Cu--Cu bond distances (in \AA) for Na$_6$Cu$_7$BiO$_4$(PO$_4$)$_4$Cl$_3$, evaluated for $U=6.66$, 7, and 8 eV. The dominant exchange paths discussed are labeled in Fig.~\ref{Na6Cu7_exchanges}. This table extends Table~\ref{table:J_for_Na6Cu7_1} by listing the full hierarchy of exchange couplings, including weaker longer-range terms.}
\label{table:J_for_Na6Cu7_3} \\
\hline\hline
$U$ (eV) & $J_1$ (K) & $J_2$ & $J_3$ & $J_4$ & $J_5$ & $J_6$ & $J_7$ & $J_8$ & $J_9$ & $J_{10}$ \\
\hline
\endfirsthead

\caption[]{(continued)} \\
\hline\hline
$U$ (eV) & $J_1$ (K) & $J_2$ & $J_3$ & $J_4$ & $J_5$ & $J_6$ & $J_7$ & $J_8$ & $J_9$ & $J_{10}$ \\
\hline
\endhead

\hline
\multicolumn{11}{r}{\footnotesize Continued on next page} \\
\endfoot

\hline\hline
\endlastfoot

6.66 & 157.7 & 202.8 & 87.3 & 29.4 & 9.0 & 0.2 & 0.0 & 1.1 & 3.9 & 47.8 \\
7    & 150.0 & 193.0 & 83.0 & 27.9 & 8.6 & 0.2 & 0.0 & 1.0 & 3.7 & 45.5 \\
8    & 131.3 & 168.8 & 72.7 & 24.4 & 7.5 & 0.2 & 0.0 & 0.9 & 3.3 & 39.8 \\
\hline
$d$ (\AA) & 3.114 & 3.278 & 4.404 & 5.009 & 5.266 & 5.340 & 5.614 & 5.727 & 5.769 & 6.018 \\
\hline
$U$ (eV) & $J_{11}$ & $J_{12}$ & $J_{13}$ & $J_{14}$ & $J_{15}$ & $J_{16}$ & $J_{17}$ & $J_{18}$ & $J_{19}$ & $J_{20}$ \\
\hline
6.66 & 0.4 & 0.1 & 1.7 & 5.0 & 0.1 & 0.0 & 0.3 & 0.0 & 0.0 & 0.1 \\
7    & 0.3 & 0.1 & 1.6 & 4.8 & 0.1 & 0.0 & 0.3 & 0.0 & 0.0 & 0.1 \\
8    & 0.3 & 0.1 & 1.4 & 4.2 & 0.1 & 0.0 & 0.2 & 0.0 & 0.0 & 0.1 \\
\hline
$d$ (\AA) & 6.234 & 6.381 & 6.555 & 7.084 & 7.717 & 7.761 & 7.805 & 7.807 & 8.120 & 8.228 \\
\end{longtable}

\begin{longtable*}{@{\extracolsep{\fill}} c c c c c c c c c @{}}
\caption{Complete set of Dzyaloshinskii--Moriya vectors $\mathbf{D}_{mn}$ (in K) for Na$_6$Cu$_7$BiO$_4$(PO$_4$)$_4$Cl$_3$ at $U=6.66$ eV. For each Cu--Cu bond, the table lists the bond distance, bond label, bond vector $\mathbf{R}_{mn}$, and the corresponding DM vector $\mathbf{D}_{mn}$. The Cu-site labels refer to Fig.~\ref{Na6Cu7_exchanges}. This table provides the full symmetry-resolved DM dataset underlying the representative values summarized in Table~\ref{table:D_for_Na6Cu7_1bond}.}
\label{table:D_for_Na6Cu7} \\
\hline\hline
$d$ (\AA) & Bond m-n & \textbf{R$_{mn}$} & \textbf{D$_{mn}$} (K) & & $d$ (\AA) & Bond m-n & \textbf{R$_{mn}$} & \textbf{D$_{mn}$} (K)\\
\hline
\endfirsthead

\caption[]{(continued)} \\
\hline\hline
$d$ (\AA) & Bond m-n & \textbf{R$_{mn}$} & \textbf{D$_{mn}$} & & $d$ (\AA) & Bond m-n & \textbf{R$_{mn}$} & \textbf{D$_{mn}$} \\
\hline
\endhead

\hline
\multicolumn{9}{r}{\footnotesize Continued on next page} \\
\endfoot

\hline\hline
\endlastfoot

3.114	&	Cu7-Cu11	&	(	-2.2	;	2.2	;	0.0	)	&	(	5.8	;	5.8	;	-11.2	)	&	&	6.018	&	Cu1-Cu8	&	(	2.8	;	5.0	;	1.8	)	&	(	0.9	;	1.7	;	2.8	)	\\
3.114	&	Cu13-Cu7	&	(	2.2	;	2.2	;	0.0	)	&	(	5.8	;	-5.8	;	-11.2	)	&	&	6.018	&	Cu1-Cu12	&	(	5.0	;	2.8	;	1.8	)	&	(	-1.7	;	-0.9	;	-2.8	)	\\
3.114	&	Cu9-Cu13	&	(	2.2	;	-2.2	;	0.0	)	&	(	-5.8	;	-5.8	;	-11.2	)	&	&	6.018	&	Cu1-Cu10	&	(	-2.8	;	5.0	;	1.8	)	&	(	0.9	;	-1.7	;	-2.8	)	\\
3.114	&	Cu11-Cu9	&	(	-2.2	;	-2.2	;	0.0	)	&	(	-5.8	;	5.8	;	-11.2	)	&	&	6.018	&	Cu1-Cu14	&	(	5.0	;	-2.8	;	1.8	)	&	(	1.7	;	-0.9	;	2.8	)	\\
3.114	&	Cu8-Cu12	&	(	2.2	;	-2.2	;	0.0	)	&	(	5.8	;	5.8	;	-11.2	)	&	&	6.234	&	Cu7-Cu4	&	(	5.3	;	-2.5	;	-2.1	)	&	(	0.0	;	-0.1	;	0.0	)	\\
3.114	&	Cu14-Cu8	&	(	-2.2	;	-2.2	;	0.0	)	&	(	5.8	;	-5.8	;	-11.2	)	&	&	6.234	&	Cu7-Cu5	&	(	5.3	;	2.5	;	-2.1	)	&	(	0.0	;	-0.1	;	0.0	)	\\
3.114	&	Cu10-Cu14	&	(	-2.2	;	2.2	;	0.0	)	&	(	-5.8	;	-5.8	;	-11.2	)	&	&	6.234	&	Cu14-Cu5	&	(	2.5	;	5.3	;	2.1	)	&	(	-0.1	;	0.0	;	0.0	)	\\
3.114	&	Cu12-Cu10	&	(	2.2	;	2.2	;	0.0	)	&	(	-5.8	;	5.8	;	-11.2	)	&	&	6.234	&	Cu14-Cu3	&	(	-2.5	;	5.3	;	2.1	)	&	(	-0.1	;	0.0	;	0.0	)	\\
3.278	&	Cu3-Cu11	&	(	-2.5	;	-0.3	;	2.1	)	&	(	7.3	;	-19.1	;	-18.8	)	&	&	6.234	&	Cu9-Cu3	&	(	-5.3	;	2.5	;	-2.1	)	&	(	0.0	;	0.1	;	0.0	)	\\
3.278	&	Cu7-Cu3	&	(	-0.3	;	-2.5	;	2.1	)	&	(	-19.1	;	7.3	;	-18.8	)	&	&	6.234	&	Cu9-Cu6	&	(	-5.3	;	-2.5	;	-2.1	)	&	(	0.0	;	0.1	;	0.0	)	\\
3.278	&	Cu11-Cu5	&	(	2.5	;	-0.3	;	2.1	)	&	(	-7.3	;	-19.1	;	-18.8	)	&	&	6.234	&	Cu12-Cu6	&	(	-2.5	;	-5.3	;	2.1	)	&	(	0.1	;	0.0	;	0.0	)	\\
3.278	&	Cu5-Cu9	&	(	0.3	;	-2.5	;	2.1	)	&	(	19.1	;	7.3	;	-18.8	)	&	&	6.234	&	Cu12-Cu4	&	(	2.5	;	-5.3	;	2.1	)	&	(	0.1	;	0.0	;	0.0	)	\\
3.278	&	Cu9-Cu4	&	(	0.3	;	2.5	;	2.1	)	&	(	19.1	;	-7.3	;	-18.8	)	&	&	6.234	&	Cu11-Cu6	&	(	2.5	;	5.3	;	-2.1	)	&	(	0.1	;	0.0	;	0.0	)	\\
3.278	&	Cu4-Cu13	&	(	2.5	;	0.3	;	2.1	)	&	(	-7.3	;	19.1	;	-18.8	)	&	&	6.234	&	Cu11-Cu4	&	(	-2.5	;	5.3	;	-2.1	)	&	(	0.1	;	0.0	;	0.0	)	\\
3.278	&	Cu13-Cu6	&	(	-2.5	;	0.3	;	2.1	)	&	(	7.3	;	19.1	;	-18.8	)	&	&	6.234	&	Cu8-Cu4	&	(	-5.3	;	2.5	;	2.1	)	&	(	0.0	;	-0.1	;	0.0	)	\\
3.278	&	Cu6-Cu7	&	(	-0.3	;	2.5	;	2.1	)	&	(	-19.1	;	-7.3	;	-18.8	)	&	&	6.234	&	Cu8-Cu5	&	(	-5.3	;	-2.5	;	2.1	)	&	(	0.0	;	-0.1	;	0.0	)	\\
3.278	&	Cu8-Cu3	&	(	0.3	;	2.5	;	-2.1	)	&	(	-19.1	;	7.3	;	-18.8	)	&	&	6.234	&	Cu5-Cu13	&	(	2.5	;	5.3	;	2.1	)	&	(	0.1	;	0.0	;	0.0	)	\\
3.278	&	Cu3-Cu12	&	(	2.5	;	0.3	;	-2.1	)	&	(	7.3	;	-19.1	;	-18.8	)	&	&	6.234	&	Cu3-Cu13	&	(	-2.5	;	5.3	;	2.1	)	&	(	0.1	;	0.0	;	0.0	)	\\
3.278	&	Cu12-Cu5	&	(	-2.5	;	0.3	;	-2.1	)	&	(	-7.3	;	-19.1	;	-18.8	)	&	&	6.234	&	Cu10-Cu3	&	(	5.3	;	-2.5	;	2.1	)	&	(	0.0	;	0.1	;	0.0	)	\\
3.278	&	Cu5-Cu10	&	(	-0.3	;	2.5	;	-2.1	)	&	(	19.1	;	7.3	;	-18.8	)	&	&	6.234	&	Cu10-Cu6	&	(	5.3	;	2.5	;	2.1	)	&	(	0.0	;	0.1	;	0.0	)	\\
3.278	&	Cu10-Cu4	&	(	-0.3	;	-2.5	;	-2.1	)	&	(	19.1	;	-7.3	;	-18.8	)	&	&	6.381	&	Cu1-Cu7	&	(	2.2	;	0.0	;	6.0	)	&	(	0.0	;	0.0	;	0.0	)	\\
3.278	&	Cu4-Cu14	&	(	-2.5	;	-0.3	;	-2.1	)	&	(	-7.3	;	19.1	;	-18.8	)	&	&	6.381	&	Cu1-Cu11	&	(	0.0	;	2.2	;	6.0	)	&	(	0.0	;	0.0	;	0.0	)	\\
3.278	&	Cu14-Cu6	&	(	2.5	;	-0.3	;	-2.1	)	&	(	7.3	;	19.1	;	-18.8	)	&	&	6.381	&	Cu1-Cu9	&	(	-2.2	;	0.0	;	6.0	)	&	(	0.0	;	0.0	;	0.0	)	\\
3.278	&	Cu6-Cu8	&	(	0.3	;	-2.5	;	-2.1	)	&	(	-19.1	;	-7.3	;	-18.8	)	&	&	6.381	&	Cu1-Cu13	&	(	0.0	;	-2.2	;	6.0	)	&	(	0.0	;	0.0	;	0.0	)	\\
4.404	&	Cu11-Cu13	&	(	0.0	;	-4.4	;	0.0	)	&	(	8.9	;	0.0	;	0.0	)	&	&	6.381	&	Cu2-Cu8	&	(	-2.2	;	0.0	;	-6.0	)	&	(	0.0	;	0.0	;	0.0	)	\\
4.404	&	Cu9-Cu7	&	(	4.4	;	0.0	;	0.0	)	&	(	0.0	;	8.9	;	0.0	)	&	&	6.381	&	Cu2-Cu10	&	(	2.2	;	0.0	;	-6.0	)	&	(	0.0	;	0.0	;	0.0	)	\\
4.404	&	Cu12-Cu14	&	(	0.0	;	4.4	;	0.0	)	&	(	8.9	;	0.0	;	0.0	)	&	&	6.381	&	Cu2-Cu12	&	(	0.0	;	-2.2	;	-6.0	)	&	(	0.0	;	0.0	;	0.0	)	\\
4.404	&	Cu10-Cu8	&	(	-4.4	;	0.0	;	0.0	)	&	(	0.0	;	8.9	;	0.0	)	&	&	6.381	&	Cu2-Cu14	&	(	0.0	;	2.2	;	-6.0	)	&	(	0.0	;	0.0	;	0.0	)	\\
5.009	&	Cu3-Cu5	&	(	-5.0	;	0.0	;	0.0	)	&	(	0.0	;	-10.8	;	1.9	)	&	&	6.555	&	Cu9-Cu10	&	(	-0.6	;	5.0	;	-4.2	)	&	(	0.0	;	0.0	;	0.0	)	\\
5.009	&	Cu5-Cu4	&	(	0.0	;	-5.0	;	0.0	)	&	(	10.8	;	0.0	;	1.9	)	&	&	6.555	&	Cu13-Cu14	&	(	5.0	;	-0.6	;	-4.2	)	&	(	0.0	;	0.0	;	0.0	)	\\
5.009	&	Cu4-Cu6	&	(	5.0	;	0.0	;	0.0	)	&	(	0.0	;	10.8	;	1.9	)	&	&	6.555	&	Cu7-Cu8	&	(	0.6	;	5.0	;	-4.2	)	&	(	0.0	;	0.0	;	0.0	)	\\
5.009	&	Cu6-Cu3	&	(	0.0	;	5.0	;	0.0	)	&	(	-10.8	;	0.0	;	1.9	)	&	&	6.555	&	Cu11-Cu12	&	(	5.0	;	0.6	;	-4.2	)	&	(	0.0	;	0.0	;	0.0	)	\\
5.266	&	Cu1-Cu4	&	(	-2.5	;	-2.5	;	3.9	)	&	(	1.3	;	-1.3	;	0.0	)	&	&	7.084	&	Cu4-Cu3	&	(	5.0	;	5.0	;	0.0	)	&	(	-1.9	;	1.9	;	0.0	)	\\
5.266	&	Cu1-Cu3	&	(	2.5	;	2.5	;	3.9	)	&	(	-1.3	;	1.3	;	0.0	)	&	&	7.084	&	Cu6-Cu5	&	(	-5.0	;	5.0	;	0.0	)	&	(	-1.9	;	-1.9	;	0.0	)	\\
5.266	&	Cu1-Cu5	&	(	-2.5	;	2.5	;	3.9	)	&	(	-1.3	;	-1.3	;	0.0	)	&	&	7.717	&	Cu1-Cu2	&	(	5.0	;	5.0	;	-3.1	)	&	(	0.0	;	0.0	;	0.0	)	\\
5.266	&	Cu1-Cu6	&	(	2.5	;	-2.5	;	3.9	)	&	(	1.3	;	1.3	;	0.0	)	&	&	7.761	&	Cu11-Cu8	&	(	2.8	;	2.8	;	6.7	)	&	(	0.0	;	0.0	;	0.0	)	\\
5.266	&	Cu2-Cu3	&	(	-2.5	;	-2.5	;	-3.9	)	&	(	-1.3	;	1.3	;	0.0	)	&	&	7.761	&	Cu11-Cu10	&	(	-2.8	;	2.8	;	6.7	)	&	(	0.0	;	0.0	;	0.0	)	\\
5.266	&	Cu2-Cu5	&	(	2.5	;	-2.5	;	-3.9	)	&	(	-1.3	;	-1.3	;	0.0	)	&	&	7.761	&	Cu13-Cu8	&	(	2.8	;	-2.8	;	6.7	)	&	(	0.0	;	0.0	;	0.0	)	\\
5.266	&	Cu2-Cu6	&	(	-2.5	;	2.5	;	-3.9	)	&	(	1.3	;	1.3	;	0.0	)	&	&	7.761	&	Cu13-Cu10	&	(	-2.8	;	-2.8	;	6.7	)	&	(	0.0	;	0.0	;	0.0	)	\\
5.266	&	Cu2-Cu4	&	(	2.5	;	2.5	;	-3.9	)	&	(	1.3	;	-1.3	;	0.0	)	&	&	7.761	&	Cu7-Cu12	&	(	2.8	;	2.8	;	6.7	)	&	(	0.0	;	0.0	;	0.0	)	\\
5.340	&	Cu1-Cu7	&	(	2.2	;	0.0	;	-4.9	)	&	(	0.0	;	0.0	;	0.0	)	&	&	7.761	&	Cu7-Cu14	&	(	2.8	;	-2.8	;	6.7	)	&	(	0.0	;	0.0	;	0.0	)	\\
5.340	&	Cu1-Cu9	&	(	-2.2	;	0.0	;	-4.9	)	&	(	0.0	;	0.0	;	0.0	)	&	&	7.761	&	Cu9-Cu12	&	(	-2.8	;	2.8	;	6.7	)	&	(	0.0	;	0.0	;	0.0	)	\\
5.340	&	Cu1-Cu11	&	(	0.0	;	2.2	;	-4.9	)	&	(	0.0	;	0.0	;	0.0	)	&	&	7.761	&	Cu9-Cu14	&	(	-2.8	;	-2.8	;	6.7	)	&	(	0.0	;	0.0	;	0.0	)	\\
5.340	&	Cu1-Cu13	&	(	0.0	;	-2.2	;	-4.9	)	&	(	0.0	;	0.0	;	0.0	)	&	&	7.805	&	Cu4-Cu14	&	(	7.5	;	-0.3	;	-2.1	)	&	(	0.0	;	0.1	;	0.0	)	\\
5.340	&	Cu2-Cu8	&	(	-2.2	;	0.0	;	4.9	)	&	(	0.0	;	0.0	;	0.0	)	&	&	7.805	&	Cu4-Cu10	&	(	-0.3	;	7.5	;	-2.1	)	&	(	-0.1	;	0.0	;	0.0	)	\\
5.340	&	Cu2-Cu10	&	(	2.2	;	0.0	;	4.9	)	&	(	0.0	;	0.0	;	0.0	)	&	&	7.805	&	Cu4-Cu13	&	(	-7.5	;	0.3	;	2.1	)	&	(	0.0	;	0.1	;	0.0	)	\\
5.340	&	Cu2-Cu12	&	(	0.0	;	-2.2	;	4.9	)	&	(	0.0	;	0.0	;	0.0	)	&	&	7.805	&	Cu4-Cu9	&	(	0.3	;	-7.5	;	2.1	)	&	(	-0.1	;	0.0	;	0.0	)	\\
5.340	&	Cu2-Cu14	&	(	0.0	;	2.2	;	4.9	)	&	(	0.0	;	0.0	;	0.0	)	&	&	7.805	&	Cu6-Cu8	&	(	0.3	;	7.5	;	-2.1	)	&	(	-0.1	;	0.0	;	0.0	)	\\
5.614	&	Cu11-Cu13	&	(	0.0	;	5.6	;	0.0	)	&	(	0.0	;	0.0	;	0.0	)	&	&	7.805	&	Cu6-Cu13	&	(	7.5	;	0.3	;	2.1	)	&	(	0.0	;	-0.1	;	0.0	)	\\
5.614	&	Cu12-Cu14	&	(	0.0	;	-5.6	;	0.0	)	&	(	0.0	;	0.0	;	0.0	)	&	&	7.805	&	Cu6-Cu14	&	(	-7.5	;	-0.3	;	-2.1	)	&	(	0.0	;	-0.1	;	0.0	)	\\
5.614	&	Cu8-Cu10	&	(	-5.6	;	0.0	;	0.0	)	&	(	0.0	;	0.0	;	0.0	)	&	&	7.805	&	Cu6-Cu7	&	(	-0.3	;	-7.5	;	2.1	)	&	(	-0.1	;	0.0	;	0.0	)	\\
5.614	&	Cu7-Cu9	&	(	5.6	;	0.0	;	0.0	)	&	(	0.0	;	0.0	;	0.0	)	&	&	7.805	&	Cu3-Cu11	&	(	7.5	;	-0.3	;	2.1	)	&	(	0.0	;	-0.1	;	0.0	)	\\
5.727	&	Cu9-Cu6	&	(	4.7	;	-2.5	;	-2.1	)	&	(	-0.2	;	0.6	;	0.5	)	&	&	7.805	&	Cu3-Cu12	&	(	-7.5	;	0.3	;	-2.1	)	&	(	0.0	;	-0.1	;	0.0	)	\\
5.727	&	Cu9-Cu3	&	(	4.7	;	2.5	;	-2.1	)	&	(	0.2	;	0.6	;	-0.5	)	&	&	7.805	&	Cu3-Cu7	&	(	-0.3	;	7.5	;	2.1	)	&	(	0.1	;	0.0	;	0.0	)	\\
5.727	&	Cu5-Cu13	&	(	2.5	;	-4.7	;	2.1	)	&	(	0.6	;	-0.2	;	0.5	)	&	&	7.805	&	Cu3-Cu8	&	(	0.3	;	-7.5	;	-2.1	)	&	(	0.1	;	0.0	;	0.0	)	\\
5.727	&	Cu13-Cu3	&	(	2.5	;	4.7	;	-2.1	)	&	(	-0.6	;	-0.2	;	0.5	)	&	&	7.805	&	Cu5-Cu12	&	(	7.5	;	0.3	;	-2.1	)	&	(	0.0	;	0.1	;	0.0	)	\\
5.727	&	Cu7-Cu4	&	(	-4.7	;	-2.5	;	-2.1	)	&	(	-0.2	;	-0.6	;	-0.5	)	&	&	7.805	&	Cu5-Cu11	&	(	-7.5	;	-0.3	;	2.1	)	&	(	0.0	;	0.1	;	0.0	)	\\
5.727	&	Cu7-Cu5	&	(	-4.7	;	2.5	;	-2.1	)	&	(	0.2	;	-0.6	;	0.5	)	&	&	7.805	&	Cu5-Cu9	&	(	0.3	;	7.5	;	2.1	)	&	(	0.1	;	0.0	;	0.0	)	\\
5.727	&	Cu11-Cu4	&	(	-2.5	;	-4.7	;	-2.1	)	&	(	0.6	;	0.2	;	0.5	)	&	&	7.805	&	Cu5-Cu10	&	(	-0.3	;	-7.5	;	-2.1	)	&	(	0.1	;	0.0	;	0.0	)	\\
5.727	&	Cu11-Cu6	&	(	2.5	;	-4.7	;	-2.1	)	&	(	0.6	;	-0.2	;	-0.5	)	&	&	7.807	&	Cu1-Cu3	&	(	2.5	;	2.5	;	-7.0	)	&	(	0.0	;	0.0	;	0.0	)	\\
5.727	&	Cu8-Cu5	&	(	4.7	;	-2.5	;	2.1	)	&	(	0.2	;	-0.6	;	0.5	)	&	&	7.807	&	Cu1-Cu4	&	(	-2.5	;	-2.5	;	-7.0	)	&	(	0.0	;	0.0	;	0.0	)	\\
5.727	&	Cu8-Cu4	&	(	4.7	;	2.5	;	2.1	)	&	(	-0.2	;	-0.6	;	-0.5	)	&	&	7.807	&	Cu1-Cu5	&	(	-2.5	;	2.5	;	-7.0	)	&	(	0.0	;	0.0	;	0.0	)	\\
5.727	&	Cu12-Cu4	&	(	2.5	;	4.7	;	2.1	)	&	(	0.6	;	0.2	;	0.5	)	&	&	7.807	&	Cu1-Cu6	&	(	2.5	;	-2.5	;	-7.0	)	&	(	0.0	;	0.0	;	0.0	)	\\
5.727	&	Cu12-Cu6	&	(	-2.5	;	4.7	;	2.1	)	&	(	0.6	;	-0.2	;	-0.5	)	&	&	7.807	&	Cu2-Cu3	&	(	-2.5	;	-2.5	;	7.0	)	&	(	0.0	;	0.0	;	0.0	)	\\
5.727	&	Cu10-Cu3	&	(	-4.7	;	-2.5	;	2.1	)	&	(	0.2	;	0.6	;	-0.5	)	&	&	7.807	&	Cu2-Cu4	&	(	2.5	;	2.5	;	7.0	)	&	(	0.0	;	0.0	;	0.0	)	\\
5.727	&	Cu10-Cu6	&	(	-4.7	;	2.5	;	2.1	)	&	(	-0.2	;	0.6	;	0.5	)	&	&	7.807	&	Cu2-Cu5	&	(	2.5	;	-2.5	;	7.0	)	&	(	0.0	;	0.0	;	0.0	)	\\
5.727	&	Cu14-Cu3	&	(	-2.5	;	-4.7	;	2.1	)	&	(	-0.6	;	-0.2	;	0.5	)	&	&	7.807	&	Cu2-Cu6	&	(	-2.5	;	2.5	;	7.0	)	&	(	0.0	;	0.0	;	0.0	)	\\
5.727	&	Cu14-Cu5	&	(	2.5	;	-4.7	;	2.1	)	&	(	-0.6	;	0.2	;	-0.5	)	&	&	8.120	&	Cu7-Cu11	&	(	7.8	;	2.2	;	0.0	)	&	(	0.0	;	0.0	;	0.0	)	\\
5.769	&	Cu7-Cu14	&	(	2.8	;	-2.8	;	-4.2	)	&	(	0.1	;	-0.1	;	0.7	)	&	&	8.120	&	Cu11-Cu9	&	(	-2.2	;	7.8	;	0.0	)	&	(	0.0	;	0.0	;	0.0	)	\\
5.769	&	Cu14-Cu9	&	(	2.8	;	2.8	;	4.2	)	&	(	-0.1	;	-0.1	;	0.7	)	&	&	8.120	&	Cu9-Cu13	&	(	2.2	;	7.8	;	0.0	)	&	(	0.0	;	0.0	;	0.0	)	\\
5.769	&	Cu9-Cu12	&	(	-2.8	;	2.8	;	-4.2	)	&	(	-0.1	;	0.1	;	0.7	)	&	&	8.120	&	Cu13-Cu7	&	(	2.2	;	-7.8	;	0.0	)	&	(	0.0	;	0.0	;	0.0	)	\\
5.769	&	Cu12-Cu7	&	(	-2.8	;	-2.8	;	4.2	)	&	(	0.1	;	0.1	;	0.7	)	&	&	8.120	&	Cu10-Cu14	&	(	7.8	;	2.2	;	0.0	)	&	(	0.0	;	0.0	;	0.0	)	\\
5.769	&	Cu10-Cu11	&	(	2.8	;	-2.8	;	4.2	)	&	(	-0.1	;	0.1	;	0.7	)	&	&	8.120	&	Cu14-Cu8	&	(	-2.2	;	7.8	;	0.0	)	&	(	0.0	;	0.0	;	0.0	)	\\
5.769	&	Cu11-Cu8	&	(	2.8	;	2.8	;	-4.2	)	&	(	0.1	;	0.1	;	0.7	)	&	&	8.120	&	Cu8-Cu12	&	(	-7.8	;	-2.2	;	0.0	)	&	(	0.0	;	0.0	;	0.0	)	\\
5.769	&	Cu8-Cu13	&	(	-2.8	;	2.8	;	4.2	)	&	(	0.1	;	-0.1	;	0.7	)	&	&	8.120	&	Cu12-Cu10	&	(	2.2	;	-7.8	;	0.0	)	&	(	0.0	;	0.0	;	0.0	)	\\
5.769	&	Cu13-Cu10	&	(	-2.8	;	-2.8	;	-4.2	)	&	(	-0.1	;	-0.1	;	0.7	)	&	&	8.228	&	Cu8-Cu9	&	(	-5.0	;	-5.0	;	4.2	)	&	(	0.0	;	0.0	;	0.0	)	\\
6.018	&	Cu2-Cu11	&	(	-5.0	;	-2.8	;	-1.8	)	&	(	-1.7	;	-0.9	;	-2.8	)	&	&	8.228	&	Cu7-Cu10	&	(	5.0	;	5.0	;	-4.2	)	&	(	0.0	;	0.0	;	0.0	)	\\
6.018	&	Cu2-Cu7	&	(	-2.8	;	-5.0	;	-1.8	)	&	(	0.9	;	1.7	;	2.8	)	&	&	8.228	&	Cu13-Cu12	&	(	5.0	;	5.0	;	-4.2	)	&	(	0.0	;	0.0	;	0.0	)	\\
6.018	&	Cu2-Cu9	&	(	2.8	;	-5.0	;	-1.8	)	&	(	0.9	;	-1.7	;	-2.8	)	&	&	8.228	&	Cu11-Cu14	&	(	5.0	;	5.0	;	-4.2	)	&	(	0.0	;	0.0	;	0.0	)	\\
6.018	&	Cu2-Cu13	&	(	-5.0	;	2.8	;	-1.8	)	&	(	1.7	;	-0.9	;	2.8	)	&	&	&		&								&		
\end{longtable*}
\twocolumngrid 


\begin{table}[h!]
    \centering
    \caption{Magnitudes $|\mathbf{D}|$ (in K) of the Dzyaloshinskii--Moriya vectors for Na$_6$Cu$_7$BiO$_4$(PO$_4$)$_4$Cl$_3$ at $U=6.66$ eV, grouped by Cu--Cu bond distance (in \AA). This condensed summary complements Table~\ref{table:D_for_Na6Cu7} by highlighting the distance dependence of the DM scale and the rapid suppression of the anisotropic exchange on longer bonds.}
    \begin{ruledtabular}
    \begin{tabular}{cccc}
	Bond distance (\AA)	&	$|$\textbf{D}$|$ & bond distance (\AA)	&	$|$\textbf{D}$|$	\\	\midrule
	3.114	&	13.9 &	6.234   &   0.2 \\
	3.278	&	27.7 &	6.381	&	0.0	\\
	4.404	&	8.9	&	6.555 & 0.0 \\	
	5.009	&	11.0 &	7.084	&	2.7	\\
	5.266	&	1.9	&	7.717	&	0.0	\\
	5.340	&	0.0	&	7.761	&	0.0	\\
	5.614	&	0.0	&	7.805	&	0.1	\\	
	5.727	&	0.8	&	7.807	&	0.0	\\	
	5.769	&	0.7	&	8.120	&	0.0	\\	
	6.018	&	3.3	&	8.228	&	0.0	\\	
    \end{tabular}
    \end{ruledtabular}
    \label{table:abs_D_for_Na6Cu7}
\end{table}


\begin{thebibliography}{49}
\bibitem [{not()}]{note_peaks}%
  \BibitemOpen
  \href@noop {} {}\bibinfo {note} {Additional subdominant Bragg peaks, carrying
  (i) 12\% of the spectral weight of the dominant ones are located at $(2\pi -
  q, q)$, $(q, 2\pi - q)$, $(-(2\pi - q), 2q)$ and $(2q, -(2\pi - q))$ and (ii)
  carrying 4\% intensity at $(-(2\pi - q), 2\pi - q)$, $(q, 2 q)$ and $(2 q,
  q)$.}\BibitemShut {Stop}%
\end{thebibliography}

%

\end{document}